\documentclass[preprint2]{aastex}

\usepackage{aas_macros}

\newcommand{\Coralie}{{\it Coralie}}
\newcommand{\Hermes}{{\it Hermes}}
\newcommand{\Hamilton}{{\it Hamilton}}
\newcommand{\kms}{km\,s$^{-1}$}
\newcommand{\ms}{m\,s$^{-1}$}
\newcommand{\asini}{$a\sin{i}$}
\newcommand{\Porb}{$P_{\rm{orb}}$}
\newcommand{\phipuls}{$\phi_{\rm{puls}}$}
\newcommand{\phiorb}{$\phi_{\rm{orb}}$}
\newcommand{\Ppuls}{$P_{\rm{puls}}$}
\newcommand{\Msol}{$M_\odot$}

\slugcomment{to appear in ApJS}

\shorttitle{Vetting Galactic Leavitt Law Calibrators}
\shortauthors{R.I.~Anderson et al.}

\listofchanges

\begin{document}


%
\title{Vetting Galactic Leavitt Law Calibrators using Radial Velocities:\\
On the Variability, Binarity, and Possible Parallax Error of 19 Long-period
Cepheids}

\author{R.~I.~Anderson\altaffilmark{1,2}, S.~Casertano\altaffilmark{3},
A.~G.~Riess\altaffilmark{1,3}, C.~Melis\altaffilmark{4},\\
B.~Holl\altaffilmark{5}, T.~Semaan\altaffilmark{5},
P.~I.~Papics\altaffilmark{6}, S.~Blanco-Cuaresma\altaffilmark{5},
L.~Eyer\altaffilmark{5}, N.~Mowlavi\altaffilmark{5}, 
L.~Palaversa\altaffilmark{5}, M.~Roelens\altaffilmark{5}}
\affil{$^1$Department of Physics and Astronomy, The Johns Hopkins University,
3400 N Charles St, Baltimore, MD 21218, USA\\
$^2$Swiss National Science Foundation Fellow\\
$^3$Space Telescope Science Institute, 3700 San Martin Dr, Baltimore, MD 21218,
USA\\
$^4$Center for Astrophysics and Space Sciences, University of California, San
Diego, CA 92093, USA\\
$^5$D\'epartement d'Astronomie, Universit\'e de Gen\`eve, 51 Ch. des Maillettes,
1290 Sauverny, Switzerland\\
$^6$Instituut voor Sterrenkunde, KU Leuven, Celestijnenlaan 200D, B-3001 Leuven,
Belgium}
\email{ria@jhu.edu}

\begin{abstract}
 We investigate the radial velocity (RV)
  variability and spectroscopic binarity of 19 Galactic long-period
  (\Ppuls $\gtrsim 10$\,d) classical Cepheid variable stars whose
  trigonometric parallaxes are being measured using the {\it Hubble}
  Space Telescope and {\it Gaia}. Our primary objective is to
  constrain possible parallax error due to undetected orbital
  motion. Using $>1600$ high-precision RVs measured between 2011 and
  2016, we find no indication of orbital motion on $\lesssim 5$\,yr
  timescales for 18 Cepheids and determine upper limits on allowed
  configurations for a range of input orbital periods. The
  results constrain the unsigned parallax error due to orbital motion
  to $< 2\,\%$ for 16 stars, and $< 4\,\%$ for 18. We improve the
  orbital solution of the known binary YZ Carinae and show that the
  astrometric model must take into account orbital motion to avoid
  significant error ($\sim \pm 100\,\mu$arcsec). We further investigate
  long-timescale (\Porb$> 10$\,yr) variations in pulsation-averaged
  velocity $v_\gamma$ via a template fitting approach using both new
  and literature RVs. We discover the spectroscopic binarity of
  XZ\,Car and CD\,Cyg, find first tentative evidence for AQ\,Car, and
  reveal KN\,Cen's orbital signature. Further (mostly tentative)
  evidence of time-variable $v_\gamma$ is found for SS\,CMa, VY\,Car,
  SZ\,Cyg, and X\,Pup. We briefly discuss considerations
  regarding a vetting process of Galactic Leavitt law calibrators and
  show that light contributions by companions are
  insignificant for most distance scale applications.
\end{abstract}


\section{Introduction}

The Cepheid\footnote{We here use the term Cepheid to denote classical type-I
Cepheids whose prototype is $\delta$\,Cephei} period-luminosity relation
\citep[PLR,][also referred to as Leavitt law]{1912HarCi.173....1L} has been a
crucial tool for determining extragalactic distances for more than a century
\citep{1913AN....196..201H}. Thanks to space-based astrometric measurements made
by the {\it Hipparcos} satellite \citep{1997ESASP1200.....P,2007ASSL..350.....V}
and the {\it Hubble} Space Telescope ({\it HST}) \citep{2002AJ....124.1695B},
this calibration has been established using absolute magnitudes estimated based
on trigonometric parallax
\citep[e.g.][]{1997MNRAS.286L...1F,2007AJ....133.1810B,2007MNRAS.379..723V}.
These Cepheid parallax measurements have greatly contributed to the overall
increase in accuracy of the determination of the local value of the Hubble
constant $H_0$
\citep{2001ApJ...553...47F,2009ApJ...699..539R,2011ApJ...730..119R}, which has recently been measured to within $2.4\%$ accuracy
\citep{2016arXiv160401424R}. Further extensive efforts are under way to reduce
this uncertainty to $1\%$ in order to improve the ability to interpret the
Cosmic Microwave Background measured using {\it PLANCK} and {\it WMAP} and learn
about the nature of Dark Energy \citep[see discussions in
e.g.][]{2012arXiv1202.4459S,2013PhR...530...87W}.

Parallax measurements are the ``gold standard'' of distance measurement, since
the technique is insensitive to the intricacies of stellar physics.
The ongoing ESA space mission {\it Gaia} is currently measuring the positions,
proper motions, and parallaxes of 1 billion stars in the Galaxy, among which
will be thousands of Cepheids \citep[and references
therein]{2012Ap+SS.341..207E}, a couple hundreds of which are expected to have
parallax determined to better than $3\%$. In the meantime,
\citet{2014ApJ...785..161R} have developed a new method of determining parallax
by spatially scanning {\it HST/WFC3}. The {\it SH0ES} team is now applying this
new technique to 19 long-period (\Ppuls$ \gtrsim 10$\,d) Galactic Cepheids,
which are particularly important for extragalactic applications of the Leavitt law, and has recently
been shown to yield the intended accuracy of $\sim 20 \hbox{--} 40 \mu$arcsec
\citep{2016ApJ...825...11C}. {\it HST/WFC3} spatial scan parallaxes provide an
important complement to {\it Gaia} parallaxes due to different systematic
uncertainties involved in narrow and wide-angle astrometry. Moreover, the {\it
HST/WFC3} parallax measurements are expected to be available before the final
{\it Gaia} data release and can anchor a new determination of $H_0$.

The imminent era of highly accurate parallaxes for hundreds of Cepheids will
enable an improved sample selection for the calibration of the Cepheid Leavitt
law. In analogy to selections made on the sample of type Ia supernovae, subsets
of Cepheids may be selected for PLR calibration depending on properties besides
fractional parallax uncertainty.
Some sample selection criteria seem obvious, for instance that objects with low
reddening are preferred or that long-period Cepheids are better analogues for
extragalactic work due to their higher luminosities.
Another potentially important point is binarity, which has been frequently
mentioned in the literature as representing a difficulty for PLR calibration.
A more complete list of considerations should include differences in the
selection and measurement process among Galactic and extragalactic Cepheids,
such as the impact of photometric zero-points, which is crucial for reducing
covariance among the various rungs of the distance ladder
\citep{2016arXiv160401424R}.
Possible differences in selection procedures include binarity and outlier
rejection.

While Galactic Cepheids may be scrutinized for binarity,
obtaining the same information for extragalactic Cepheids does not currently seem feasible.
Similarly, Galactic Cepheids offer the opportunity to study Cepheid variability
in greater detail than extragalactic Cepheids.
Historically, the concept of stellar populations introduced by 
\citet{1944ApJ...100..137B} eventually resulted in the understanding that
type-II and type-I Cepheids follow different PLRs and had a tremendous impact on
the understanding of the size and age of the universe  \citep[for a discussion,
see][]{1956PASP...68....5B}.
More subtle differences may yet exist among the objects now classified as type-I
Cepheids, and detailed studies of Galactic members of this class will be
essential for investigating this possibility.

Taking a first step towards clarifying the role of binarity on PLR calibration,  we here present a detailed investigation of \emph{spectroscopic} binarity of the 19 Cepheids for which {\it HST/WFC3} spatial scan parallaxes are being recorded. The primary aims of this investigation are to take stock of the spectroscopic binarity of
the program stars, as well as to set upper limits on undetected companions and
the potential parallax error resulting from modeling the {\it HST} astrometric
data of a binary Cepheid as a single star.
We further provide a detailed description of the morphology of the radial
velocity (RV) variability of the program Cepheids, report the average
velocities, and briefly consider the potential for unresolved companions
stars to affect Cepheid luminosity estimation.

This paper is structured as follows. \S\ref{sec:selection} presents
the initial selection of the program stars. The following
\S\ref{sec:obs} describes more than $1600$ high-precision RV
observations obtained using three telescopes and spectrographs.
\S\ref{sec:modeling} describes the modeling of RV curves for pulsation
and orbital motion. \S\ref{sec:RVvariability} presents the pulsational
variability of the program Cepheids.  \S\ref{sec:caveats} discusses
caveats involved in such modeling of high-precision Cepheid RV
data. \S\ref{sec:binarity} presents the results obtained related to
spectroscopic binarity. In \S\ref{sec:upperlims} we determine upper
limits on undetected RV orbital motion over the baseline of our
observations (\Porb$\lesssim 5\,$yr) and use these results to
constrain possible parallax error
due to orbital motion for 18 of the 19 program Cepheids.
\S\ref{sec:YZCar} presents an improved orbital solution for
YZ\,Carinae as well as an estimation of this orbit's influence on the
parallax measurement. Longer-timescale (\Porb $> 10$\,yr)
spectroscopic binarity is investigated in
\S\ref{sec:LongTimeScaleOrbits}, which is divided into subsections for
newly-reported candidates (\S\ref{sec:newbinaries}) and ones
previously discussed in the literature (\S\ref{sec:oldbinaries})
Additional considerations pertaining to the (general) binarity and
variability of Cepheids in the context of distance measurements are
provided in \S\ref{sec:discussion}. The final \S\ref{sec:conclusions}
summarizes all results.

\section{Observations \& Data}

\subsection{Sample selection}\label{sec:selection}
The Cepheid sample investigated here was selected according to several criteria.
As the primary goal is to determine parallax accurately using {\it HST/WFC3}
spatial scans \citep[see][]{2014ApJ...785..161R,2016ApJ...825...11C}, the most
crucial selection criteria were ones centered on the astrometric measurement
itself.

An optimal target for high-accuracy spatial scan parallax measurements has 
\begin{itemize}
  \item pulsation period longer than approximately 10\,days as this reflects the
  periods of the predominant group of Cepheids found in other galaxies (due to higher
  luminosity) (S.~L.~Hoffmann et al., submitted) and avoids putative
  non-linearities of the PLR intervening at $10$\,d
  \cite[for differing recent opinions on the matter, see][]{2013ApJ...764...84I,2016MNRAS.457.1644B,2016ApJ...824...74G}
  \item mean $V-$band magnitude fainter than $7.5$ to avoid saturation ($H-$band
  $> 5$\,mag).
  \item at least 5, ideally more than 10, reference stars within $6$ magnitudes
  of the Cepheid for which trails would be recorded simultaneously.
  \item an expected distance of $d \lesssim 3$\,kpc so that parallax can be
  determined to better than $10\%$ for each individual Cepheid for a nominal parallax
  uncertainty of $20 \hbox{--} 40\,\mu$arcsec.
  \item no known companion star with \Porb\ on the order of the
  sparsely-sampled {\it HST/WFC3} spatial scan observations (typically 5 epochs). 
  \item extinction ($A_H \lesssim 0.5$\,mag) to avoid excessive uncertainty in
  the inferred absolute magnitudes.
\end{itemize}

Although binaries were not strictly excluded from the sample
(see, e.g., YZ\,Carinae below or visual binaries), we
caution that the present sample is subject to selection effects concerning binarity
and should not be considered random in this regard. Therefore, we stress that
this sample alone should not be used to infer the properties of binary fractions
unless this occurs in conjunction with further observational data capable of
eliminating or reducing such selection effects \citep{2013AJ....146...93R}.

\subsection{Description of observations}\label{sec:obs}

\floattable
\begin{deluxetable}{lrrrcrrrrr}
\tablecaption{Basic information on the sample of Cepheids discussed here}
\tablehead{\colhead{Cepheid} & \colhead{HD} & \colhead{RA(J2000)} &
\colhead{DE(J2000)} & \colhead{$\langle m_V \rangle$} & \colhead{$N_{\rm{Cor}}$}
& \colhead{$N_{\rm{Ham}}$} & \colhead{$N_{\rm{Her}}$} & \colhead{$\Delta
t_{\rm{obs}}$} & \colhead{References} \\ 
 & \colhead{} & \colhead{[h:m:s]} & \colhead{[d:m:s]} & \colhead{[mag]} &
\colhead{} & \colhead{} & \colhead{} & \colhead{[yr]} & \colhead{} }
\startdata
SY Aur  & 277622 & 05:12:39.20 & 42:49:54 & 9.1 & 0 & 78 & 31 & 2.6 &  $-$ \\ 
SS CMa  &  HIP\,36088  & 07:26:07.20 & -25:15:26 & 9.9 & 24 & 42 & 14 & 3.0 & 
1, 2, 3 \\
VY Car  & 93203 & 10:44:32.70 & -57:33:55 & 7.5 & 83 & 0 & 0 & 5.0 &  1, 2, 4,
5, 6 \\
XY Car  & 308149 & 11:02:16.10 & -64:15:46 & 9.3 & 70 & 0 & 0 & 1.9 &  $-$ \\ 
XZ Car  & 305996 & 11:04:13.50 & -60:58:48 & 8.6 & 118 & 0 & 0 & 4.2 &  \S\ref{sec:newbinaries} \\
YZ Car  & 90912 & 10:28:16.80 & -59:21:01 & 8.7 & 28 & 0 & 0 & 2.4 &  7, 8, \S\ref{sec:YZCar} \\
AQ Car  & 89991 & 10:21:23.00 & -61:04:27 & 8.9 & 59 & 0 & 0 & 1.9 &  \S\ref{sec:newbinaries} \\
HW Car  & 92490 & 10:39:20.30 & -61:09:09 & 9.2 & 68 & 0 & 0 & 5.0 &  $-$ \\
DD Cas  &  HIP\,118122  & 23:57:35.00 & 62:43:06 & 9.9 & 0 & 78 & 23 & 2.3 &  9, 10 \\
KN Cen  &  HIP\,66383  & 13:36:36.90 & -64:33:30 & 9.9 & 70 & 0 & 0 & 2.1 &  9, 11, 12, 13, 14, 15 \\
SZ Cyg  & 196018 & 20:32:54.30 & 46:36:05 & 9.4 & 0 & 66 & 41 & 2.3 &  16, 17, \S\ref{sec:newbinaries} \\
CD Cyg  & 227463 & 20:04:26.60 & 34:06:44 & 9.0 & 0 & 66 & 45 & 2.3 & 
\S\ref{sec:newbinaries} \\
VX Per  & 236948 & 02:07:48.50 & 58:26:37 & 9.4 & 0 & 78 & 40 & 4.0 &  $-$ \\ 
X Pup  & 60266 & 07:32:47.00 & -20:54:35 & 8.6 & 48 & 21 & 15 & 1.4 &  18 \\ 
AQ Pup  & 65589 & 07:58:22.10 & -29:07:48 & 8.7 & 51 & 41 & 6 & 3.0 &  10, 19,
20 \\
WZ Sgr  & 167660 & 18:16:59.70 & -19:04:33 & 8.1 & 48 & 6 & 29 & 4.0 &  5, 15,
21, 22, 23 \\
RY Sco  & 162102 & 17:50:52.30 & -33:42:20 & 8.0 & 52 & 0 & 0 & 2.1 &  1, 22 \\
Z Sct  & 172902 & 18:42:57.30 & -05:49:15 & 9.6 & 41 & 48 & 41 & 3.1 &  $-$ \\ 
S Vul  & 338867 & 19:48:23.80 & 27:17:11 & 9.1 & 12 & 12 & 37 & 1.9 &  $-$ \\
\enddata
\tablecomments{Basic information on the sample of Cepheids discussed. Hipparcos
identifiers are given where no HD number was available. Coordinates and mean
magnitudes are based on the information from the
GCVS\footnote{\url{http://www.sai.msu.su/gcvs/gcvs/}}, average magnitudes are
approximate. The number of observations obtained with \Coralie, \Hamilton, and
\Hermes\ are listed together with the total temporal baseline $\Delta
t_{\rm{obs}}$ of our new observations. Typically, we obtained three observations
per pointing with the \Hamilton\ spectrograph. The total number of new
observations made available here is $1630$.} 
\tablereferences{References for work previously discussing the binarity or
cluster membership of these objects are as follows (see also the binary Cepheids database by \citet{2003IBVS.5394....1S}): 1:  \citet{1994AJ....108..653E}, 2:  \citet{1996A+A...311..189S}, 3:  \citet{2016ApJ...825...11C}, 4:  \citet{1977AJ.....82..163T}, 5:  \citet{2013MNRAS.434.2238A}, 6:  \citet{1997ESASP1200.....P}, 7:  \citet{1983MNRAS.205.1135C}, 8:  \citet{2004MNRAS.350...95P}, 9:  \citet{1977MNRAS.178..505M}, 10: \citet{1980PASP...92..315M}, 11: \citet{1964BAN....17..520W}, 12: \citet{1968MNRAS.141..109L}, 13: \citet{1970MNRAS.148....1S}, 14: \citet{1978A+A....62...75P}, 15: \citet{1989CoKon..94.....S}, 16: \citet{1966PZ.....16...10K}, 17: \citet{1991CoKon..96..123S}, 18: \citet{2012MNRAS.426.3154S}, 19: \citet{1966AJ.....71..999F}, 20: \citet{1991Ap+SS.183...17V}, 21: \citet{2002ApJS..140..465B}, 22: \citet{1981A+AS...44..179P}, 23: \citet{1993ApJS...85..119T}. Sections discussing individual Cepheids in more detail are indicated in the column Refs.}
\label{tab:SampleObs}
\end{deluxetable}

We have secured time-series observations from three different high-resolution
echelle spectrographs: \Coralie\ ($R\sim 60,000$) at the Swiss $1.2$\,m Euler
telescope located at La Silla Observatory, Chile; \Hermes\ ($R\sim85,000$) at
the Flemish $1.2$\,m Mercator
telescope\footnote{\url{http://www.mercator.iac.es/}} located at the Roque de
los Muchachos Observatory on La Palma, Canary Islands, Spain; \Hamilton\
($R\sim60,000$) at the $3$\,m Shane telescope located at Lick Observatory,
California, USA.

\Coralie\ and \Hermes\ spectra were reduced using the dedicated pipelines
available on site. \Hamilton\ spectra were reduced using standard {\tt IRAF}
routines. All spectra were bias-corrected and flat-fielded,
and cosmic ray hits were removed. ThAr (\Coralie, \Hermes) and TiAr
\citep[\Hamilton]{2013AJ....146...97P} lamps were used for wavelength
calibration. 

All radial velocities (RVs) presented here were determined using the
cross-correlation technique \citep{1996A+AS..119..373B,2002A+A...388..632P}
using a numerical mask representative of a solar spectral type (G2 mask).

\Coralie\ RVs were corrected for temporal variations in the wavelength
calibration using ThAr reference spectra that are interlaced with the science
orders on the detector. \Hermes\ spectra were corrected for such variations
using frequent re-calibration of the wavelength solution and a model for
estimating RV zero-point changes associated with changes in air pressure
\citep[as done in][]{2015ApJ...804..144A}. \Hamilton\ spectra are the most
affected by temporal variations in the instrumental zero-point, which dominate 
the uncertainty for the associated RVs presented here. We use
stable RV standard stars to track the RV variation due to intra-night changes of
the wavelength solution and determine appropriate corrections for
science exposures by interpolating the time sequence of offsets determined.

The precision of \Coralie\ and \Hermes\ measurements is typically on the order
of $10 \hbox{--} 30\,$\ms, depending on the signal-to-noise ratio (SNR)
achieved. At this level of precision, the instrumental
zero-points of \Coralie\ and \Hermes\ are compatible without adjustments.
\Hamilton\ RVs are significantly less precise due to the unstable zero-point;
here we adopt $200\,$\ms\ as a typical uncertainty for \Hamilton\ RVs. This
value includes the uncertainty associated with tracking the nightly zero-point
variations using standard stars as well as (smaller) RV zero-point differences
among instruments.

The time of observation for all newly-observed spectra are given as solar system
barycentric Julian dates minus $2\,400\,000$ and all associated RV measurements
are relative to the solar system barycenter.

\section{Radial Velocity Modeling}

\floattable
\begin{deluxetable}{lrrrrr}
\tablecaption{Example RV data obtained for this program}
\tablehead{ \colhead{Cepheid} & \colhead{BJD$- 2.4$M} & \colhead{\phipuls} &
\colhead{$v_r$} & \colhead{$\sigma(v_r)$} & \colhead{Instrument} \\
 & \colhead{[d]} & \colhead{} & \colhead{[\kms]} &
 \colhead{[\kms]} & }
\startdata 
SY Aur  &  56402.68783  &  0.0580  &  -10.248  &  0.2  &  Hamilton \\
SY Aur  &  56519.01162  &  0.5236  &  6.516  &  0.2  &  Hamilton \\
SY Aur  &  56519.01798  &  0.5242  &  6.500  &  0.2  &  Hamilton \\
SY Aur  &  56581.00590  &  0.6341  &  10.463  &  0.2  &  Hamilton \\
SY Aur  &  56581.01220  &  0.6347  &  10.463  &  0.2  &  Hamilton \\
SY Aur  &  56581.01850  &  0.6354  &  10.496  &  0.2  &  Hamilton \\
SY Aur  &  56581.91266  &  0.7235  &  7.004  &  0.2  &  Hamilton \\
SY Aur  &  56581.91897  &  0.7241  &  6.930  &  0.2  &  Hamilton \\
SY Aur  &  56581.92527  &  0.7247  &  6.837  &  0.2  &  Hamilton \\
SY Aur  &  56609.97389  &  0.4894  &  4.237  &  0.2  &  Hamilton \\
\multicolumn{6}{c}{\ldots}\\
S Vul  &  57500.899168  &  0.7406  &  15.663  &  0.032  &  Coralie \\
S Vul  &  57504.910489  &  0.7991  &  14.473  &  0.073  &  Coralie \\
S Vul  &  57507.895237  &  0.8426  &  10.209  &  0.035  &  Coralie \\
S Vul  &  57508.895744  &  0.8572  &  7.892  &  0.029  &  Coralie \\
S Vul  &  57511.898062  &  0.9010  &  -0.430  &  0.032  &  Coralie \\
S Vul  &  57526.868106  &  0.1192  &  -10.978  &  0.016  &  Coralie \\
S Vul  &  57528.902885  &  0.1489  &  -9.989  &  0.022  &  Coralie \\
S Vul  &  57529.903860  &  0.1635  &  -9.466  &  0.025  &  Coralie \\
S Vul  &  57534.837524  &  0.2105  &  -6.426  & 0.014  & Coralie \\
S Vul  &  57536.886856  &  0.2399  &  -5.069  & 0.024  & Coralie \\
\enddata
\tablecomments{The full dataset comprising $1630$ observations of the 19
Cepheids obtained with the three spectrographs is published online via the
Journal and the {\tt CDS}. Pulsation phase is defined such that
\phipuls$ = 0$ at minimal RV and is computed using ephemerides listed in
Tab.\,\ref{tab:RVresults}. Dates and RVs are relative to the Solar
system barycenter.}
\end{deluxetable}

\subsection{Methodology}\label{sec:modeling}

The observed RV curve of a (binary) Cepheid is a superposition of the systemic
RV relative to the Solar system barycenter, $v_\gamma$, the pulsational
variability, $v_{r,\rm{puls}}$, and the orbital motion of the Cepheid relative
to the center of gravity of the binary system, $v_{r,\rm{orb}}$. Thus,
\begin{equation}
v_r (t) = v_\gamma + v_{r,\rm{puls}}+ v_{r,\rm{orb}} \,.
\label{eq:sum}
\end{equation}

We model the pulsational variability as a Fourier series with an appropriate
(fixed) number of harmonics, $N_{\rm{FS}}$, which is adopted during a
preliminary inspection of the available RV data, cf. Tab.\,\ref{tab:RVresults}.
The Fourier model of the pulsation is computed as:
\begin{equation}
v_{r,\rm{puls}} (t) = \sum_{i=1}^{N}{ a_i \sin{ 2\pi \phi_{\rm{puls}}
} + b_i \cos{ 2\pi \phi_{\rm{puls}}}}\, , 
\label{eq:FSeries}
\end{equation}
with pulsation phase $\phi_{\rm{puls}} = (t - E) / P_{\rm{puls}}$, where $t$ is
time in Julian days, $E$ is the reference epoch, and \Ppuls\ is determined by minimizing
the internal scatter of our RV data using as starting point a reference
value from the General Catalog of Variable Stars \citep{2009yCat....102025S}.
We here employ a definition of the epoch $E$ so that $\phi = 0.0$ coincides
with minimal RV near the mean (solar system) barycentric JD of the data
considered.
This choice of phase zero-point is arbitrary and not of particular importance to
this work. $\phi = 0.0$ is expected to be close to a time of maximum
light although the values of $E$
presented here are not necessarily comparable to 
times of maximum light determined from light curves.
Similarly, the values of \Ppuls\ that minimize scatter of the present
RV data are close, albeit not necessarily identical, to \Ppuls\ listed in the
literature, depending on the number of harmonics used for the fit as well as
how well the pulsations repeat over time.
One exception to this procedure is the case of YZ\,Carinae, where the strong
orbital RV signal complicates the determination of \Ppuls\ based on RV data alone. We
therefore used $V$-band photometric data from the All Sky Automated Survey
\citep{2002AcA....52..397P} to derive a new best-fit \Ppuls\ and adopted this
value for the RV modeling.

In general, RV orbital motion is modeled as a Keplerian with
semi-amplitude $K$, eccentricity $e$, argument of periastron $\omega$, and the
true anomaly $\theta$ \citep[see e.g.][]{2001icbs.book.....H}:
\begin{equation}
v_{r,\rm{orb}} = K \left[ \cos{\left( \omega + \theta \right) } + e
\cos{\omega} \right] \,.
\label{eq:Keplerian} 
\end{equation}

However, the majority of Cepheids considered here do not exhibit evidence of
orbital motion, and indeed one of  the primary aims of this work is to set upper
limits on undetected companions. We thus assume zero eccentricity 
unless required (and explicitly stated). This simplifies Eq.\,\ref{eq:Keplerian}
to an ordinary sinusoid with amplitude and phase, which we here model as:
\begin{equation}
v_{r,\rm{orb,e=0}} = a_{\rm{orb}} \sin{ 2\pi \phi_{\rm{orb}}} + b_{\rm{orb}}
\cos{ 2\pi \phi_{\rm{orb}}} \, ,
\label{eq:sinusoid}
\end{equation}
with orbital semi-amplitude $K = \sqrt{a_{\rm{orb}}^2 + b_{\rm{orb}}^2}$. 
Orbital period and semi-amplitude then yield the projected semimajor axis
$a\sin{i}$ of the Cepheid's orbit around the common center of
gravity:
\begin{equation}
a \sin{i}\,\rm{[AU]} = 9.192 \times 10^{-5} \cdot K\,[\rm{km\,s^{-1}}] \cdot
P_{\rm{orb}}\,[\rm{d}] \, .
\label{eq:asini}
\end{equation}

Highly eccentric or very long orbital period (\Porb\,$\gg\,5$\,yr) systems may
remain undetected by our RV measurements, depending on the geometry and which part of
the orbit would be sampled by the observations. To this end, we also inspect the
long-term stability of $v_\gamma$ using a combination of our new data with
published RVs from the literature, see \S\ref{sec:LongTimeScaleOrbits}. However,
companions on such very long-period orbits (\Porb$>10$\,yr) are not
likely to affect the {\it HST/WFC3} parallax measurements.

In the following subsections, we discuss the pulsational RV modeling of the
program Cepheids. The search for spectroscopic binarity and estimation of
parallax error due to companions is presented in \S\ref{sec:binarity} below.

\subsection{RV variability of program Cepheids}
\label{sec:RVvariability}
\begin{figure*}
\centering
\includegraphics{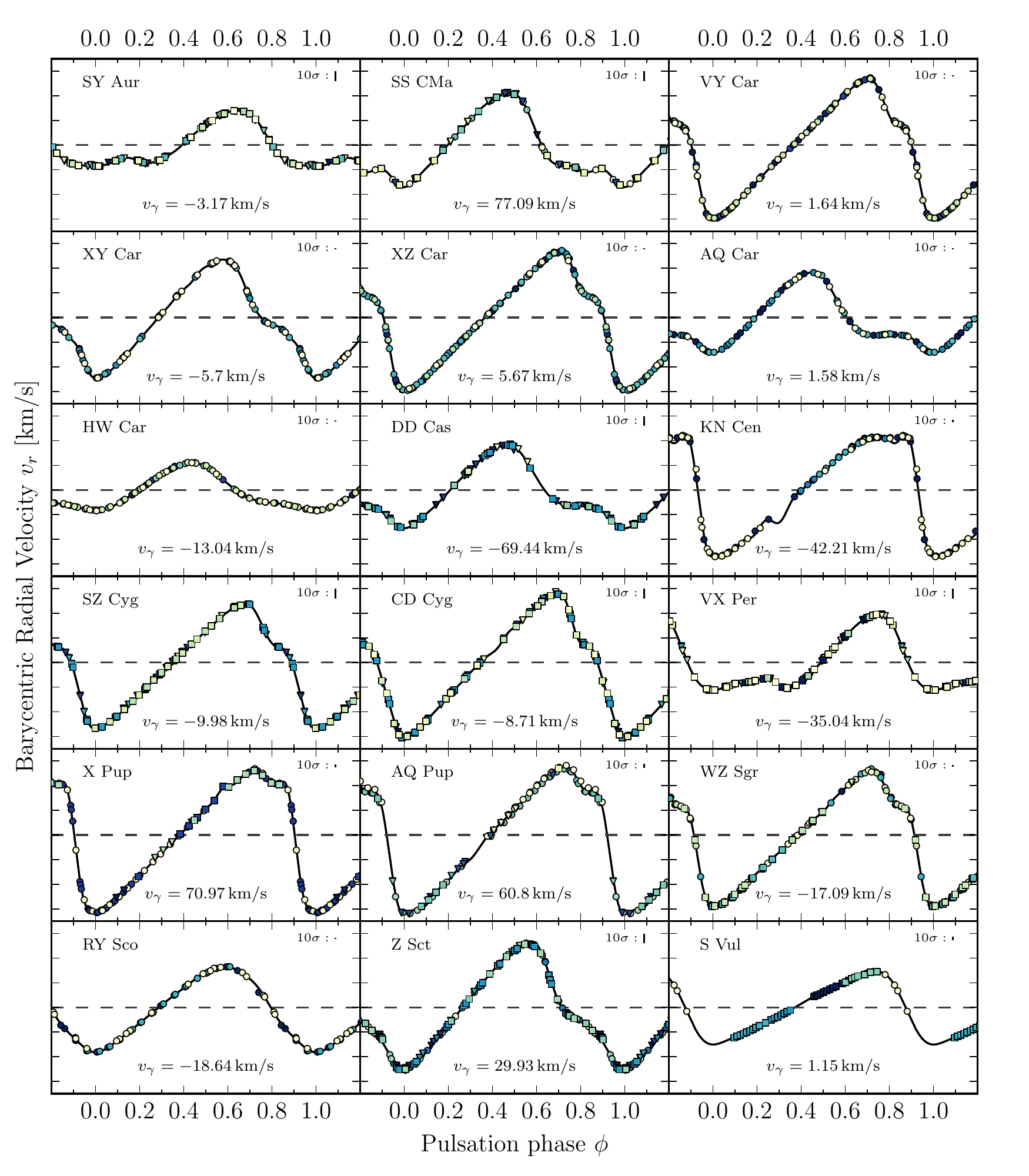}
\caption{Phase-folded new RV measurements with fitted Fourier series.
Color traces observation date increasing from blue to yellow, cf.
Fig.\,\ref{fig:OMCvsBJD}. The dashed horizontal line indicates
$v_\gamma$ and each subplot's y-range is $v_\gamma \pm
35$\,\kms. The ten-fold median uncertainty is indicated in each upper right
corner. Circles identify data from \Coralie, squares from \Hermes, and triangles
from the \Hamilton\ spectrograph. }
\label{fig:RVvsPHI}
\end{figure*}

\begin{figure*}
\centering
\includegraphics{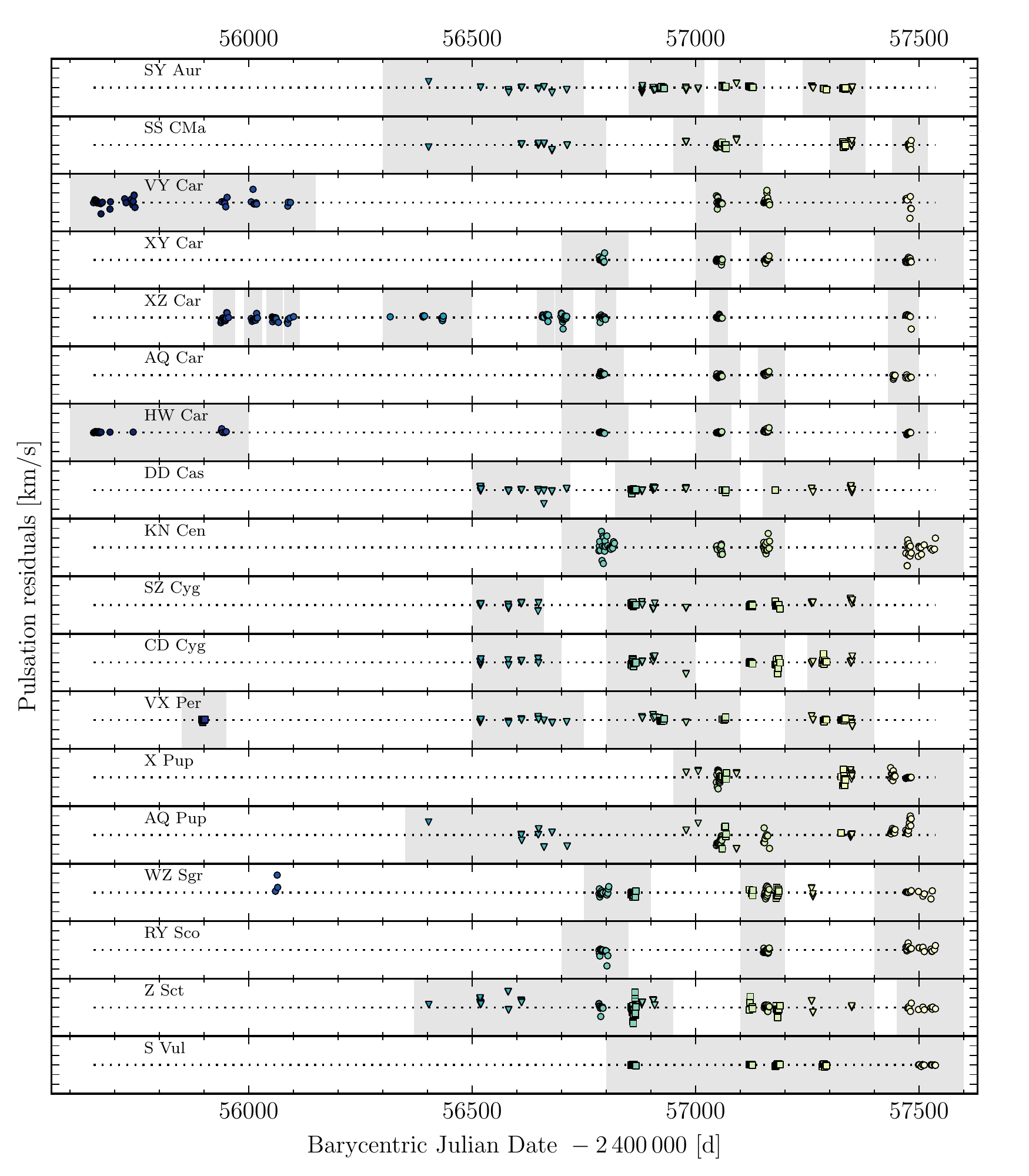}
\caption{RV residuals after pulsation modeling against (solar system)
barycentric Julian date.
The ordinate of each subplot is centered on $0\,$\kms with a range of $\pm
3$\,\kms. Date ranges (tranches) suitable for template fitting are indicated by
shaded backgrounds, cf. \S\ref{sec:LongTimeScaleOrbits}.}
\label{fig:OMCvsBJD}
\end{figure*}

Figure\,\ref{fig:RVvsPHI} presents the data for 18 of the program
stars (YZ\,Car is described separately in Sec.\,\ref{sec:YZCar}) together with
the fitted pulsation model. Figure\,\ref{fig:OMCvsBJD}
shows the corresponding residuals as a function of observation date. The figures
illustrate that most Cepheids have very well-sampled RV curves, although a few
cases could benefit from better phase-sampling. This includes in particular
S\,Vul, which is observationally challenging due to its
extremely long and unstable pulsation period
\citep{1978ATsir1003....4M,1980IBVS.1895....1M}.
We here find a best-fit period of \Ppuls$=69.7\,$d, which is approximately
$1.4\%$ longer than previously reported values, compared to the 
range of periods in the literature ($67.3 \hbox{--} 68.7$\,d). Nevertheless, our
observations do not sample the complete pulsation curve (in particular the
minimum RV), introducing a substantial systematic period uncertainty of $\sim
1$\,d.

The spectroscopic binarity of Cepheids is usually determined by investigating
the long-term stability of the pulsation-averaged velocity $v_\gamma$.
Specifically, all known Cepheid binaries have \Porb$> 1$\,yr and $K > 1\,$\kms\
\citep[cf.][]{2003IBVS.5394....1S}, possible smaller amplitude companions being
masked by RV zero-point offsets among different instruments on the order of a
few hundred \ms\ \citep[cf.][]{2015AJ....150...13E} or noise intrinsic to
Cepheid RV variability. The consistently flat residuals shown in
Fig.\,\ref{fig:OMCvsBJD} thus provide no indication for spectroscopic binarity,
leaving only the possibility of very low-amplitude ($K \lesssim 1$\,\kms) or
very long-timescale (\Porb$\gg 5$\,yr) orbital motion. These possibilities
are investigated in detail in Sec.\,\ref{sec:binarity}. 

\floattable
\begin{deluxetable}{lrrrrrrrrrrrr}
\tablecaption{Results from RV curve modeling for all program Cepheids}
\tablehead{ \colhead{Cepheid} & \colhead{\Ppuls} & \colhead{$E$} &
\colhead{$N_{\rm{FS}}$} & \colhead{$v_\gamma$} & \colhead{$A_{\rm{p2p}}$} &
\colhead{$A_1$} & \colhead{$R_{21}$} & \colhead{$\phi_{21}$} &
\colhead{$R_{31}$} & \colhead{$\phi_{31}$} & \colhead{rms} &
\colhead{$\sigma_{v_\gamma}$} \\   
 & \colhead{[d]} & \colhead{JD$-2.4$M} & \colhead{} & \colhead{[\kms]} &
 \colhead{[\kms]} & \colhead{[\kms]} & \colhead{} & \colhead{} & \colhead{} & &
 \colhead{[\kms]} & \colhead{[\kms]} }
\startdata 
SY Aur & 10.145458 & 56990.536188 & 9 & -3.172 & 22.403 & 10.524 & 0.39 & 4.33 & 0.09 & 6.99 & 0.216 & 0.051 \\ 
SS CMa & 12.352828 & 57062.269144 & 7 & 77.085 & 38.345 & 16.801 & 0.24 & 4.80 &
0.17 & 1.62 & 0.274 & 0.110 \\
VY Car & 18.882696 & 56502.102998 & 9 & 1.637 & 57.358 & 23.973 & 0.29
& 3.01 & 0.07 & 5.27 & 0.438 & 0.281 \\
XY Car & 12.436275 & 57145.616572 & 11 & -5.698 & 48.171 & 20.987 &
0.03 & 3.61 & 0.13 & 2.78 & 0.176 & 0.161 \\
XZ Car & 16.652208 & 56554.510768 & 13 & 5.671 & 56.299 & 23.788 &
0.29 & 3.01 & 0.07 & 5.22 & 0.274 & 0.150 \\
YZ Car$^*$ & 18.1676 & 51928.9358 & 8 & 0.844 & 29.692 & 14.078 & 0.08 & 3.14 &
0.04 & 2.19 & 0.037 & 0.063 \\
AQ Car & 9.769452 & 57105.87848 & 7 & 1.577 & 32.932 & 13.976 & 0.30 & 5.08 &
0.17 & 1.75 & 0.167 & 0.121 \\
HW Car & 9.199135 & 56727.566552 & 7 & -13.035 & 19.284 & 9.018 & 0.22
& 5.05 & 0.08 & 1.74 & 0.128 & 0.075 \\
DD Cas & 9.812156 & 56871.673758 & 7 & -69.453 & 33.913 & 14.486 & 0.24 & 5.24 & 0.17 & 2.01 & 0.163 & 0.039 \\ 
KN Cen & 34.018969 & 57135.280429 & 9 & -42.217 & 50.014 & 22.343 &
0.33 & 3.00 & 0.21 & 5.93 & 0.821 & 0.514 \\
SZ Cyg & 15.11133 & 56921.702556 & 9 & -9.976 & 51.060 & 21.949 & 0.24 & 2.94 &
0.06 & 4.73 & 0.255 & 0.051 \\
CD Cyg & 17.076041 & 56946.237797 & 11 & -8.709 & 58.892 & 24.724 & 0.27 & 3.01
& 0.05 & 5.00 & 0.352 & 0.083 \\
VX Per & 10.882827 & 56819.055602 & 7 & -35.037 & 30.790 & 13.471 & 0.47 & 4.45
& 0.16 & 7.41 & 0.224 & 0.051 \\
X Pup & 25.959165 & 57262.157494 & 11 & 70.970 & 57.069 & 25.403 & 0.36
& 3.00 & 0.17 & 5.77 & 0.430 & 0.162 \\
AQ Pup & 30.182036 & 57123.479592 & 9 & 60.798 & 59.522 & 26.096 &
0.32 & 2.98 & 0.14 & 5.67 & 0.777 & 0.375 \\
WZ Sgr & 21.850992 & 57052.286612 & 11 & -17.088 & 54.919 & 23.601 & 0.32 & 3.05
& 0.10 & 5.55 & 0.370 & 0.185 \\
RY Sco & 20.322084 & 57172.826932 & 7 & -18.653 & 34.768 & 16.375 & 0.16 & 2.93
& 0.01 & 4.79 & 0.330 & 0.294 \\
Z Sct & 12.901867 & 56956.219712 & 7 & 29.924 & 52.156 & 22.614 & 0.06 & 4.09 &
0.14 & 2.46 & 0.590 & 0.120 \\
S Vul & 69.653841 & 57241.56046 & 5 & 1.137 & 29.766 & 12.962 & 0.35 &
3.12 & 0.15 & 6.24 & 0.070 & 0.033 \\
\enddata 
\tablecomments{Pulsation periods and epochs of minimal RV determined for the number of
harmonics indicated ($N_{FS}$). Mean velocity $v_\gamma$, peak-to-peak amplitude $A_{\rm{p2p}}$, first harmonic amplitude $A_1$, Fourier amplitude and phase ratios $R_{21}$,
$\phi_{21}$, $R_{31}$, $\phi_{31}$, fit rms, and standard mean error on
$v_\gamma$. $^*$: YZ\,Carinae is a spectroscopic binary, cf.
Tab.\,\ref{tab:YZCarOrbit} and Sec.\,\ref{sec:YZCar}. }
\label{tab:RVresults}
\end{deluxetable}

\begin{figure*}
\centering
\includegraphics{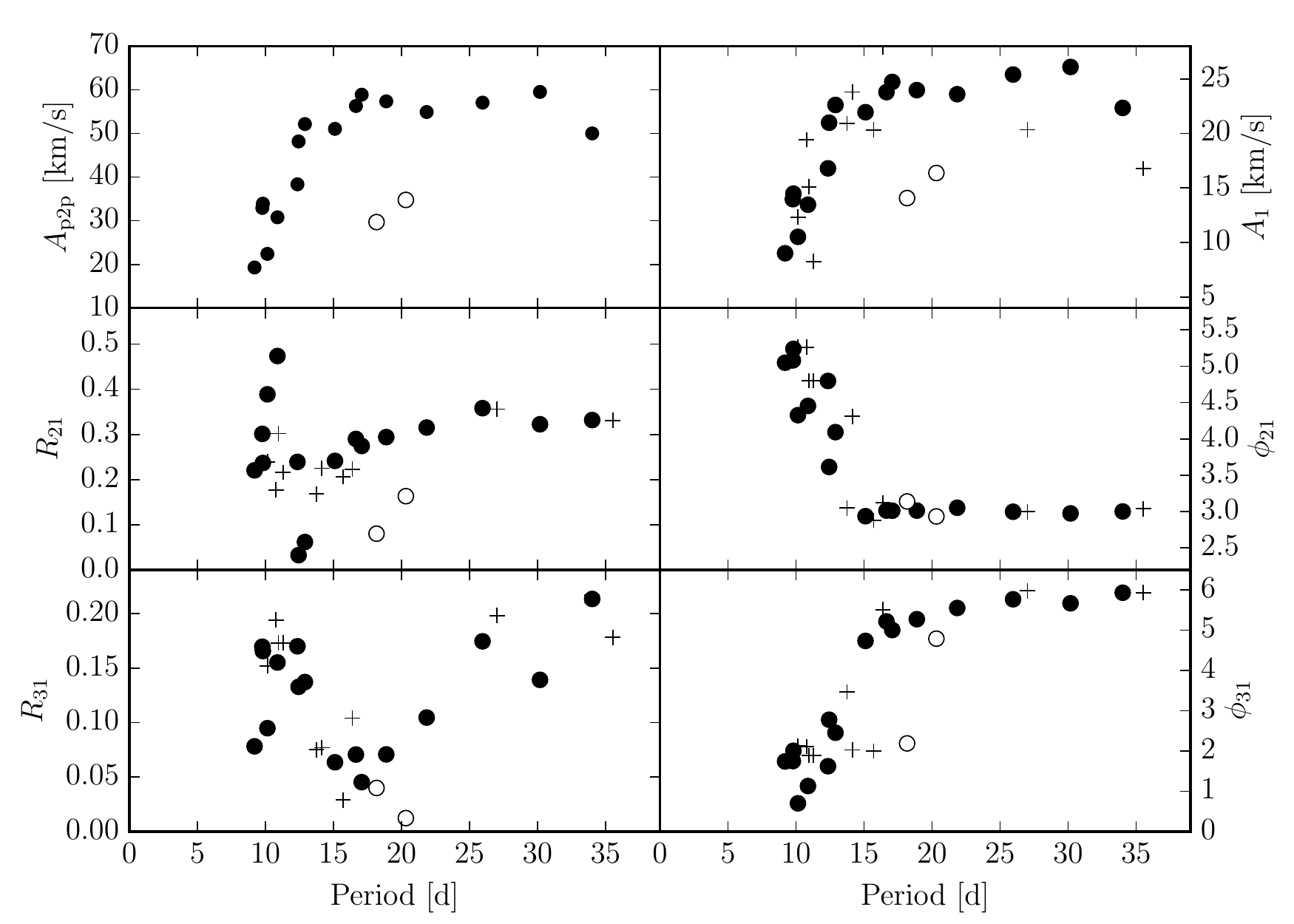}
\caption{Fourier parameters of the program Cepheids against \Ppuls. S\,Vul
(\Ppuls$\sim 69$\,d) is excluded for clarity. YZ\,Car and RY\,Sco
are marked as open circles. Pluses denote additional Cepheid parameters
published by \citet{1990ApJ...351..606K}.}
\label{fig:FourierParms}
\end{figure*}

Table\,\ref{tab:RVresults} lists the results of the pulsational modeling for all
program Cepheids. Specifically, it includes best-fit pulsation periods and
epochs, number of harmonics used for the Fourier series $N_{\rm{FS}}$, systemic velocity
$v_\gamma$ \citep[which will be of use for {\it Gaia},
see][]{2012A+A...546A..61D}, peak-to-peak amplitude $A_{\rm{p2p}}$, amplitude of
the first harmonic $A_1$, 
Fourier ratios $R_{21}$, $R_{31}$, $\phi_{21}$, and $\phi_{31}$
\citep{1981ApJ...248..291S}, fit rms and uncertainty on $v_\gamma$.
Amplitude and phase of the $i$-th harmonic are defined as $A_i = \sqrt{a_i^2 +
b_i^2}$ and $\tan{\phi_i} = b_i/a_i$, and are computed using the
coefficients obtained from Eq.\,\ref{eq:FSeries}. Amplitude ratios among
harmonics are defined as $R_{i1} = A_i / A_1$, and phase ratios as $\phi_{i1} =
\phi_i - i\cdot\phi_1$. Figure\,\ref{fig:FourierParms} illustrates these
results and their dependence on logarithmic \Ppuls, ignoring S\,Vul for which
the available RV data were insufficient to reliably determine these parameters.

We find a dependence of Fourier amplitude and phase ratios on \Ppuls\ in broad
agreement with previous observational \citep{1990ApJ...351..606K} and
simulation-based results \citep{2000A+A...363..593A}. In particular, we recover
the general morphology of increasing RV amplitudes that flatten off around 17
days, as well as the associated steep decline in $\phi_{21}$. The period
distribution of our program Cepheids nicely complements the sample presented by
\citet{1990ApJ...351..606K}, doubling the number of Cepheids in the \Ppuls\ 
range upward of 10 days. These parameters will be
useful for hydrodynamical modeling of Cepheid variability, although such
applications are outside the scope of this work. Here, we use these parameters
to show that most Cepheids exhibit the RV variability behavior expected for
their \Ppuls.

We note that the RV amplitudes of RY Sco and YZ Car are outliers from the
overall trend indicated by the other stars in the sample. Additionally, visual
inspection of the data reveals that the (pulsational part of the) RV curve is
more sinusoidal than that of other Cepheids, cf. Fig.\,\ref{fig:RVvsPHI} and
\S\,\ref{sec:YZCar} for YZ Car. This simple RV curve shape is quantified as low
amplitude ratios between the first three harmonics, cf. parameters $R_{21}$ and
$R_{31}$ in Fig.\,\ref{fig:FourierParms}. However, it is known from photometric
studies that light curve amplitudes can vary considerably at fixed \Ppuls\
according to the pulsation-average temperature of the Cepheid, i.e., its
position in the instability strip
\citep[e.g.][]{1996A+A...314..541A,2009A+A...493..471S,2010MNRAS.408..695K}.
Inspection of the {\it ASAS} lightcurves \citep{2002AcA....52..397P} of RY Sco
and YZ Car shows that both exhibit very similar, saw-tooth-shaped $V-$band
variability with only a very minor bump-like feature near minimum light and
similar peak-to-peak amplitude of $\sim 0.8$\,mag. Since other Cepheids in
this \Ppuls\ range exhibit stronger bump features in their light curves and
larger RV amplitudes, this suggests a connection between the low RV amplitudes of RY Sco and YZ Car and the weak bumps. We consider a detailed investigation of the 
connection between RV and light curve shapes to be out of the scope of this
work. In the near future, our parallax measurements will help to clarify
whether these differences in light and RV curve shapes among Cepheids are
related to differences in luminosity. 

\subsection{Caveats of Cepheid RV curve modeling}
\label{sec:Chi2}\label{sec:caveats}

Pulsation period changes due to secular evolution\hbox{---}i.e., linear
variations on the order of $10 \hbox{--} 100\,\rm{s\,yr^{-1}}$ for solar metallicity Cepheids in the period
range $10 - 50$\,d \citep[e.g.][]{2016A+A...591A...8A}\hbox{---}are generally
not an issue over the less than 5-year temporal baseline of our observations.
However, Cepheids are also known to exhibit non-linear pulsation period changes
over shorter timescales, and this effect is particularly noticeable for
\Ppuls$\gtrsim 20$\,d.
\citep[e.g.][]{1989CoKon..94.....S,1991CoKon..96..123S,2000NewA....4..625B,2009AstL...35..406B,2008AcA....58..313P,2014A+A...566L..10A}.
A well-known example is RS Pup (\Ppuls$\sim
42$\,d), whose non-linear \Ppuls\ variations can result in
phase offsets of up to $20\%$ over the course of 20 years
\citep{2009AstL...35..406B}. 
For short-period overtone Cepheids, period fluctuations on similar timescales
have been found using high-cadence photometry from {\it Kepler} and {\it MOST}
\citep{2012MNRAS.425.1312D,2015AJ....150...13E}.

The peak-to-peak RV amplitudes presented here are on the order of $20 \hbox{--}
60$\,\kms, cf. Tab.\,\ref{tab:RVresults}. Using instruments featuring extreme
long-term instrumental stability and high RV precision on the order of a few
\ms, it is now possible to detect pulsation irregularities on the order of
$0.01\%$ with confidence.
This has led to the discovery of RV curve modulation
\citep{2014A+A...566L..10A}, which is particularly erratic in long-period
(\Ppuls$>10$\,d) Cepheids, where cycle-to-cycle variations are found.
For example, RS Pup's RV amplitude varies by approximately 1\,\kms\ from one
pulsation cycle to the next, and up to 3\,\kms\ over the course of one year.
Both effects are also present, albeit weaker, in the $35$\,d Cepheid
$\ell$\,Car.

Very dense time-sampling is required to model non-linear period and
amplitude fluctuations on a cycle-to-cycle basis \citep{2016MNRAS.455.4231A}. 
Cepheids with \Ppuls $\gtrsim 20$\,d are most affected by these difficulties,
since non-linear period fluctuations are strongest for these stars and since
achieving good phase-sampling is particularly challenging due to
practical constraints such as telescope access, weather, and the Moon. 

Modeling Cepheid RV variability using the adopted stable model
(Eq.\,\ref{eq:FSeries}) thus fails to account for all (astrophysical) signals
present in the data, leading to excess residuals and generally very high values
of $\chi^2$.
Since these high $\chi^2$ values are the result of model inadequacy, it would be
incorrect to scale RV uncertainties, which have furthermore been shown to
represent an adequate estimation of RV precision in the sense of the ability to
reproduce a central value from multiple measurements
\citep{2013PhDT.......363A}. 

The presence of additional signals dominates the reduction in $\chi^2$ when a
further model component is introduced in the fit, such as orbital motion
(Eq.\ref{eq:Keplerian} or \ref{eq:sinusoid}).
We therefore caution that a detection of spectroscopic binarity should not be
claimed based purely on a reduction in $\chi^2$.
Rather, additional (visual) inspection of the data is required and must be
weighed against other indicators.

\section{Investigating spectroscopic binarity}
\label{sec:binarity}

\subsection{Constraining parallax error due to orbital motion}
\label{sec:upperlims}

\begin{figure*}
\centering
\begin{tabular}{ll}
\includegraphics{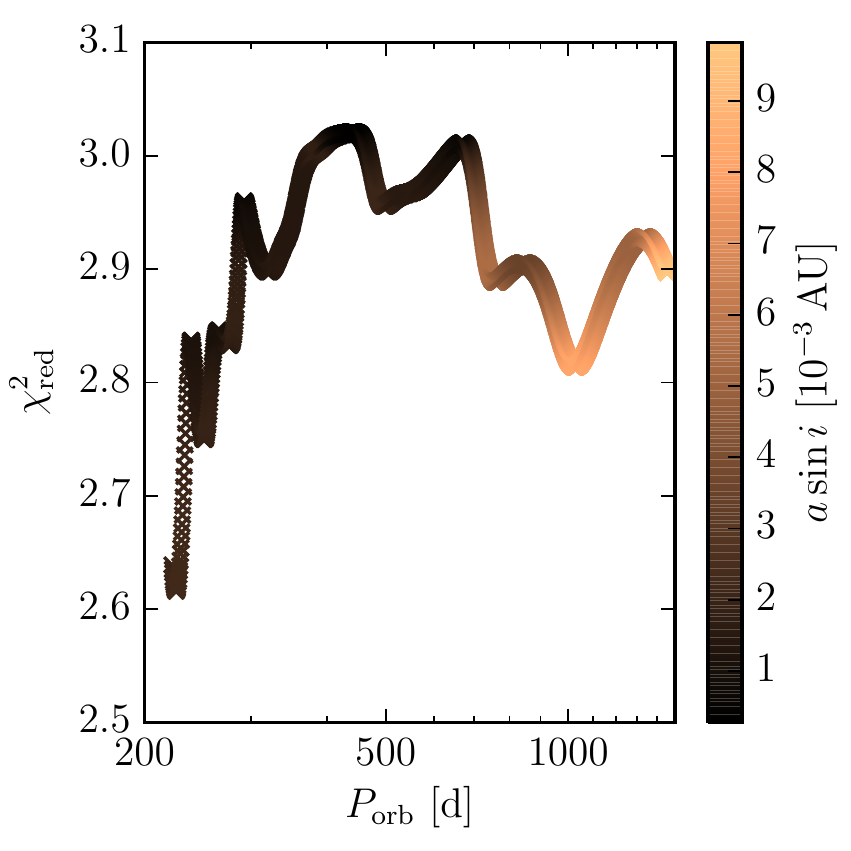} &
\includegraphics{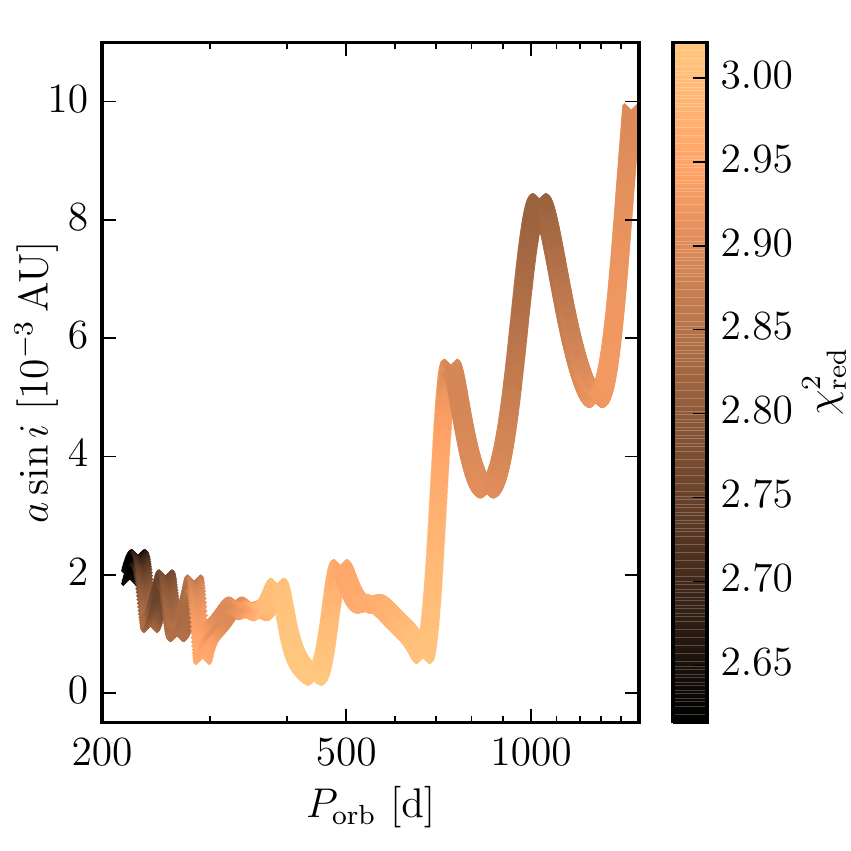}
\end{tabular}
\caption{$\chi^2$ map and best-fit projected semimajor axes ($a\sin{i}$ in [AU])
as a function of (fixed) orbital period for VX\,Persei.}
\label{fig:VXPer_a1sini_Porb}
\end{figure*}

Binarity of variable stars can affect parallax measurements primarily in two ways:
1) via \emph{real} positional modulation due to orbital motion that is not
accounted for by the astrometric modeling and 2) via \emph{apparent} positional
modulation in phase with the variability (e.g. of the Cepheid). Here we are
interested primarily in the former effect and how to constrain the related error
using RVs. Objects affected by photocenter variations due to variability were
denoted as Variability Induced Movers \citep[VIMs]{1996A+A...314..679W} in {\it
Hipparcos} and will be considered in future work, since RVs cannot constrain
this effect. It is worth noting that parallax error due to orbital
motion can be strong even for low-mass companions that would not lead to
VIM-type error, see the example of $\delta$\,Cep \citep{2015ApJ...804..144A}
whose companion is at least $5\hbox{--}6$ $H-$band magnitudes fainter than the
primary \citep{2016arXiv160601108G}.

Parallax error due to orbital motion (henceforth: parallax error) is expected to
be strongest for \Porb\ near one year due to the aliasing between orbital and
parallactic motion of the {\it HST} spatial scan observations. 
In this section we estimate the possible hidden impact of binarity on the {\it
HST} parallax measurements by determining upper limits on the potential
astrometric impact of undetected binaries among the program Cepheids.
Photometric effects are briefly discussed in \S\ref{sec:photometric}. We first
determine orbital configurations of spectroscopic binaries not ruled out by the
high-precision RV data for a range of input orbital periods covering the
range of possible binary orbits (here: \Porb $> 222$\,d) up to the temporal
baseline of the measurements\footnote{All known orbital periods of Cepheid binaries are
larger than one year, cf. \citet{2003IBVS.5394....1S,2015AJ....150...13E}}.
We then constrain the parallax error $\hat{\varpi}$ that could result from modeling the {\it
HST} astrometry of the Cepheid as a single star if in reality it were a binary.

The temporal baseline of the RV data presented here is ideal for
this purpose. The data have been recorded
contemporaneously with (within approximately one year of) the {\it HST}
spatial scan measurements and cover the range of orbital periods where the greatest impact of orbital motion on the parallax measurement is to be expected (longer \Porb\ would primarily affect proper
motion). However, while RV data are highly sensitive, they can only measure the
line-of-sight component of orbital motion. This limitation is explicitly
described in the following.

We model the RV data as a sum of mean velocity, pulsation, and circular orbital
motion (Eq.\,\ref{eq:sum} with Eq.\,\ref{eq:sinusoid}) for a set of input
(fixed) orbital periods \Porb\ using the pulsation ephemerides listed in
Tab.\,\ref{tab:RVresults}.
For each \Porb, we obtain a best-fit solution for $v_\gamma$ and the Fourier
coefficients, as well as a semi-amplitude $K$ and orbital phase \phiorb. The
projected semimajor axis of each best-fit solution is calculated using
Eq.\,\ref{eq:asini}. We use \asini\ in AU, since the angular size of the orbit
scales with the parallax of the Cepheid.

Figure\,\ref{fig:VXPer_a1sini_Porb} shows the results obtained using this
procedure for the example of VX\,Persei. The left hand panel shows the
$\chi^2$ map, which would tend to favor \Porb $< 1$\,year with
very small \asini, specified here in [AU], i.e., equivalent to the
fraction of the star's parallax.

To estimate parallax error from these upper limits on spectroscopic binarity, we
first compute positional offsets $\delta x(t)$ due to orbital motion at times
$t$ of {\it HST} spatial scan observations using the best-fit values of \asini\ and
\phiorb\ determined for each \Porb. Although RV measurements are blind to
inclination, positional offsets due to orbital motion intrinsically depend both
on inclination $i$ and the orientation of the line of nodes with respect to the
astrometric resolution direction, $\theta$. We therefore compute positional
offsets for a two-dimensional grid of inclination and orientation angles for
each orbital period \Porb\ used in the RV modeling. Each positional offset is
computed as:
\begin{equation}
\delta x(t) = a \cos{\phi} \cos{\theta} + a \sin{\phi} \sin{\theta} \cos{i}\ ,
\label{eq:posoffset}
\end{equation}
where $a = a\sin{i}/\sin{i}$ is the semimajor axis of the Cepheid around the
center of gravity of the hypothetical binary, and $\phi = 2\pi\frac{t -
E}{P_{\rm{orb}}} + \phi_{\rm{orb},0}$ with $E$ the epoch of the RV modeling (cf.
Tab.\,\ref{tab:RVresults}) and $\tan{\phi_{\rm{orb},0}} = B_{\rm{orb}} /
A_{\rm{orb}}$, cf. Eq.\,\ref{eq:sinusoid}.

Finally, we estimate the projected parallax error $\hat{\varpi}\sin{i}$
using the set of positional offsets computed for each orbital period
using a least-squares procedure that takes into account the exact
times $t$ and parallax factors $\pi_f$ of each {\it HST}
observation. Hence, we compute $\hat{\varpi}(P_{\rm{orb}},i,\theta)$
using the $\delta x(t)$ for all {\it HST} observations.  As $i$ and
$\theta$ are unconstrained by the RV data and since we are interested
in conservative upper limits, we adopt the maximal unsigned
$\hat{\varpi}$ for each \Porb\ and multiply this value by the sine of
the inclination for which it was computed, i.e.,
\begin{equation}
\hat{\varpi}\sin{i} (P_{\rm{orb}}) = \mathrm{max}( \vert \hat{\varpi} 
(P_{\mathrm{orb}},i,\theta) \vert ) \cdot \sin{i_{\mathrm{max( \vert \hat{\varpi} \vert)}}}
\label{eq:plxbias}
\end{equation}
Note that we here use the absolute value $\vert \hat{\varpi} (P_{\mathrm{orb}},i,\theta) \vert$ to estimate the unsigned projected parallax error, since RV data do not constrain $i$ and $\theta$. Depending on the configuration, $\hat{\varpi}\sin{i} (P_{\rm{orb}})$ could be positive or negative, resulting in an over- or underestimated parallax.

The quantity $\hat{\varpi}\sin{i}$ represents an upper limit in the sense
that it reflects the maximal unsigned parallax error for a given orbital period. 
However, it preserves the notion that the basis for this upper limit, the modeling of RV
data, cannot constrain inclination. We further note that $1/\sin{i} < 3$ for $i
> 19.5\deg$ ($94\%$ of possible inclinations) and $< 10$ for $i > 5.7\deg$
($99.5\%$).

Table\,\ref{tab:upperlimits} lists the results thus obtained for all
program Cepheids excluding YZ\,Car, whose orbit is updated in
\S\ref{sec:YZCar} below.  For each Cepheid, we provide information for
1) the solution offering the weakest constraint
  on possible parallax error, and 2) for the solution with minimal
$\chi^2$.

As mentioned in \S\ref{sec:RVvariability} above, we find no indication
for binarity for these 18 Cepheids for a range of orbital periods on
the order of the observational baseline ($2\hbox{--}5$\,yr, depending
on the star).  More importantly, the minimum-$\chi^2$ solutions
provide an estimated mean upper limit on parallax
error of $0.8\%$ for all 18 stars ($< 1\%$ for 13,
$< 4\%$ for all 18), despite some imperfections in the sampling of
some Cepheids. We caution however that the systematic uncertainty of
\Ppuls\ due to incomplete phase coverage may affect the result for
S\,Vul, cf. \S\ref{sec:RVvariability}, although additional
observations are required to determine whether this is the case.

The stars with the weakest constraints on $\max{(\vert
  \hat{\varpi}\sin{i} \vert)}$ are RY\,Sco ($7\%$), AQ\,Pup
($6.7\%$), X\,Pup ($4.2\%$), and KN\,Cen ($3.1\%$), all of which
exhibit signs of cycle-to-cycle fluctuations of pulsation period
and/or amplitude. The 14 remaining Cepheids have $\max{(\hat{\varpi}\sin{i})} < 2\,\%$, even for these solution with
maximal impact. As expected, most of the best-fit orbital solutions
that would lead to the greatest parallax error are near the 1\,yr alias
between orbital and parallactic motion. Both exceptions for which this
is not the case, SY\,Aur and VX\,Per, yield the largest parallax error
at \Porb\ corresponding to the baseline of the measurements.  These
results therefore strongly suggest that an astrometric modeling
assuming a single star configuration is appropriate for all 18
Cepheids, see \S\ref{sec:YZCar} for the exception of YZ\,Car.

Qualitatively, the flat pulsation-only residuals shown in
Fig.\,\ref{fig:OMCvsBJD} already indicated that no large parallax error due to
orbital motion was to be expected for these stars. The above results for \asini\
and parallax error mirror and \emph{quantify} this point.
While a main limitation of this RV-based work is its insensitivity to
inclination, this quantification of possible undetected configurations serves to
increase confidence in the accuracy of the parallax measurements themselves and
will be useful for future vetting of candidate high-accuracy calibrators of the
Galactic Leavitt law.

\floattable
\begin{deluxetable}{lcc|rrrrr|rrrrr}
\rotate
\tablecaption{Upper limits on $\hat{\varpi} \sin{i}$, the maximum unsigned 
projected parallax error estimated from upper limits on
spectroscopic binarity} 
\tablehead{ & & & \multicolumn{5}{c}{greatest impact on parallax} &  
\multicolumn{5}{c}{overall minimum $\chi^2$} \\
\colhead{Cepheid} & \colhead{$\Delta t$} & \colhead{DOF} &
\colhead{\Porb} & \colhead{$K$} & \colhead{\asini} &
\colhead{$\hat{\varpi}\sin{i}$} & \colhead{$\chi^2$} & 
\colhead{\Porb} & \colhead{$K$} & \colhead{\asini} &
\colhead{$\hat{\varpi}\sin{i}$} & \colhead{$\chi^2$} \\
 & \colhead{[yr]} &  & [yr] & \colhead{[\kms]} & \colhead{[$\%\ \varpi$]}  &
 \colhead{[$\%\ \varpi$]} & & \colhead{[yr]} & \colhead{[\kms]} &
 \colhead{[$\%\ \varpi$]} &
 \colhead{[$\%\ \varpi$]} & }
\startdata
SY Aur & 2.6 & 87 & 2.6 & 0.186 $\pm$ 0.001 & 1.63 & 0.0 & 125 & 2.07 & 0.179 $\pm$ 0.001 & 1.24 & 0.0 & 107 \\
SS CMa & 2.96 & 62 & 0.76 & 0.191 $\pm$ 0.022 & 0.49 & 0.2 & 867 & 0.77 & 0.193 $\pm$ 0.021 & 0.5 & 0.2 & 866 \\ 
VY Car & 5.01 & 61 & 0.95 & 0.242 $\pm$ 0.028 & 0.78 & 0.9 & 20618 & 0.64 & 0.103 $\pm$ 0.008 & 0.22 & 0.1 & 20207 \\ 
XY Car & 1.91 & 44 & 1.07 & 0.421 $\pm$ 0.025 & 1.52 & 1.7 & 3238 & 1.09 & 0.43 $\pm$ 0.029 & 1.58 & 1.6 & 3219 \\ 
XZ Car & 4.23 & 88 & 0.99 & 0.185 $\pm$ 0.005 & 0.61 & 0.8 & 10928 & 2.48 & 0.197 $\pm$ 0.001 & 1.64 & 0.6 & 8556 \\ 
AQ Car & 1.91 & 41 & 1.0 & 0.476 $\pm$ 0.012 & 1.6 & 1.9 & 709 & 1.01 & 0.472 $\pm$ 0.018 & 1.59 & 1.9 & 708 \\ 
HW Car & 5.01 & 50 & 0.96 & 0.149 $\pm$ 0.003 & 0.48 & 0.6 & 789 & 0.61 & 0.108 $\pm$ 0.001 & 0.22 & 0.1 & 544 \\
DD Cas & 2.28 & 82 & 1.05 & 0.099 $\pm$ 0.001 & 0.35 & 0.4 & 81 & 1.11 & 0.1 $\pm$ 0.001 & 0.37 & 0.4 & 80 \\ 
KN Cen & 2.06 & 47 & 0.97 & 0.858 $\pm$ 0.568 & 2.81 & 3.1 & 19831 & 0.99 & 0.763 $\pm$ 0.724 & 2.54 & 2.8 & 19791 \\ 
SZ Cyg & 2.28 & 85 & 0.96 & 0.076 $\pm$ 0.001 & 0.24 & 0.3 & 234 & 1.75 & 0.147 $\pm$ 0.001 & 0.87 & 0.1 & 204 \\ 
CD Cyg & 2.28 & 85 & 1.06 & 0.122 $\pm$ 0.002 & 0.43 & 0.6 & 552 & 1.96 & 0.142 $\pm$ 0.002 & 0.93 & 0.2 & 529 \\ 
VX Per & 3.99 & 100 & 3.99 & 0.073 $\pm$ 0.002 & 0.98 & 0.2 & 290 & 0.62 & 0.105 $\pm$ 0.001 & 0.22 & 0.0 & 262 \\ 
X Pup & 1.38 & 58 & 1.13 & 1.168 $\pm$ 0.098 & 4.43 & 4.2 & 2093 & 1.14 & 1.128 $\pm$ 0.093 & 4.31 & 3.9 & 2077 \\ 
AQ Pup & 2.96 & 76 & 1.01 & 1.567 $\pm$ 0.047 & 5.32 & 6.7 & 12079 & 1.66 &
1.038 $\pm$ 0.004 & 5.77 & 0.4 & 2299 \\ 
WZ Sgr & 4.02 & 57 & 0.93 & 0.222 $\pm$ 0.075 & 0.69 & 0.7 & 5126 & 1.9 & 0.406 $\pm$ 0.023 & 2.59 & 0.6 & 4913 \\ 
RY Sco & 2.06 & 33 & 0.99 & 1.901 $\pm$ 1.161 & 6.34 & 7.0 & 13759 & 0.67 & 0.41 $\pm$ 0.007 & 0.93 & 0.2 & 13424 \\ 
Z Sct & 3.1 & 111 & 0.94 & 0.444 $\pm$ 0.019 & 1.4 & 1.5 & 3156 & 0.93 & 0.45 $\pm$ 0.018 & 1.41 & 1.4 & 3148 \\ 
S Vul & 1.87 & 34 & 0.97 & 0.075 $\pm$ 0.005 & 0.24 & 0.3 & 32 & 0.61 & 0.192 $\pm$ 0.001 & 0.39 & 0.1 & 27 \\
\enddata
\tablecomments{For each star, we list the temporal baseline of our new observations,
the degrees of freedom after the fit, as well as best-fit results for a) the
orbital solution leading to the maximum unsigned projected parallax error (typically
near 1 year orbital period), and b) the orbital solution with overall minimal
$\chi^2$ (see Sec.\,\ref{sec:Chi2} for a related discussion).
Uncertainties on $K$ are based on the fit covariance matrix. \asini\
is calculated using $K$ and the corresponding fixed \Porb\ (cf. Eq.\,\ref{eq:asini}).
$\hat{\varpi}\sin{i}$ is estimated as described in \S\ref{sec:upperlims} and can have positive or negative sign. $\hat{\varpi}\sin{i}$ and \asini\ are given in [AU],
which is equivalent to percent of the parallax. For all solutions shown here, eccentricity $e = 0$.
The orbital elements of YZ\,Carinae are given separately in
Tab.\,\ref{tab:YZCarOrbit}.}
\label{tab:upperlimits}
\end{deluxetable}

\subsection{Updating the orbital solution of YZ\,Carinae}\label{sec:YZCar} 

YZ\,Carinae (\Ppuls $= 18.1676$\d) is the only  
spectroscopic binary Cepheid among the program stars whose \Porb\ is shorter
than our observational baseline.
Its spectroscopic binary nature was discovered and originally reported by
\citet{1983MNRAS.205.1135C} together with a preliminary orbital estimate of \Porb\ $\sim 850$\,d and low eccentricity.
\citet{2004MNRAS.350...95P} obtained additional, higher-precision RV data and
determined a significantly shorter orbital period of $657$\,d with similar
eccentricity.

\begin{figure*}
\centering
\includegraphics{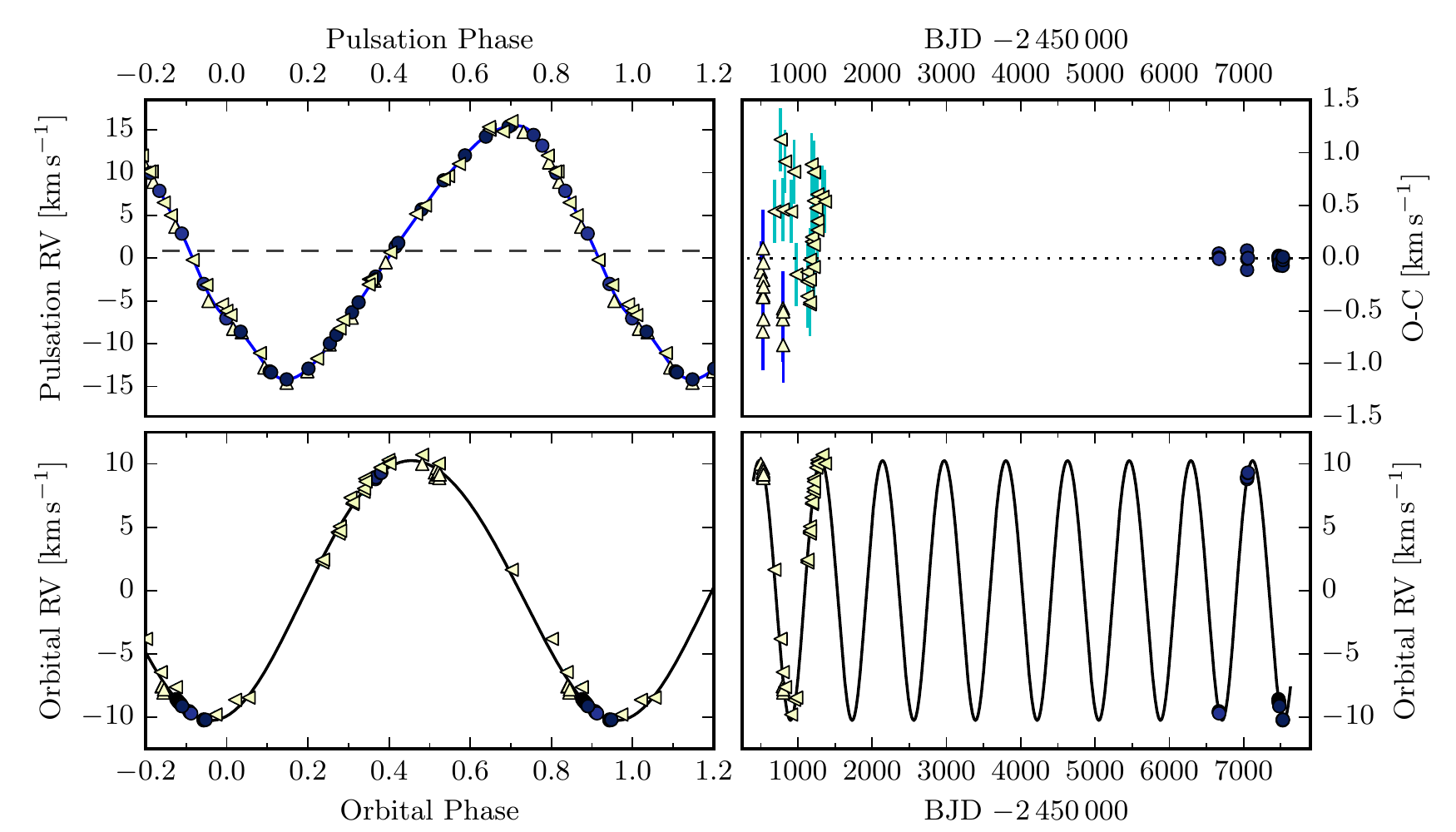}
\caption{Orbit of YZ Carinae. Color and symbols are the same in all panels. From
top left to bottom right: Pulsation-only RV variability incl. systemic velocity
$0.844$\,\kms indicated by horizontal dashed line; residuals of data
minus Fourier series and Keplerian orbit versus observation date; orbital motion
phase-folded with orbital period; orbital motion versus observation date.
Yellow upward triangles are from \citet{2002ApJS..140..465B}, yellow leftward
triangles from \citet{2004MNRAS.350...95P}. Blue circles are \Coralie\
measurements.}
\label{fig:YZCar_Orbit}
\end{figure*}

We here update and improve YZ\,Car's orbital solution by fitting a combination
of a Fourier series and a Keplerian orbit to the new, highly precise,
\Coralie\ data presented here together with RVs published by \cite{2002ApJS..140..465B} and the
post-1996 measurements by \citet[Tab.\,A4]{2004MNRAS.350...95P}.
Figure\,\ref{fig:YZCar_Orbit} illustrates the quality of this solution.  
To verify our result, we also determined the orbit including older 
measurements by \cite{1994A+AS..105..165P} and \citet{1983MNRAS.205.1135C} in
the fit, finding excellent agreement. However, these older
data do not improve the quality of the solution due to larger measurement
uncertainties and/or the possibility of pulsation period changes and we
therefore prefer the solution based exclusively on RVs with uncertainties better
than $300$\,\ms.
The value of \Ppuls\ adopted for this modeling was determined using {\it ASAS}
$V-$band photometry, since orbital motion significantly affects the measured RV on timescales of a month. Details of
YZ\,Car's orbital solution and previous determinations are provided in
Tab.\,\ref{tab:YZCarOrbit}. 

The value of \Porb$\sim 830$\,d determined here is nearly in agreement with the
rough estimate provided by \citet[$850$\,d]{1983MNRAS.205.1135C}, and strongly
disagrees with the solution presented by
\citet[$657.3$\,d]{2004MNRAS.350...95P}, which is striking due to the small
uncertainties quoted in the latter publication.
Based on a visual inspection of the various available data sets, we conclude that the pre-1996 data employed by
\citet[Tab.\,A3]{2004MNRAS.350...95P} (pre-1996 MJUO RVs) are not comparable
with the other available RV data. This mismatch of data could be explained by
the fact that the pre-1996 MJUO RVs were measured by 
different collaborators who may have employed nonstandard definitions of RV,
such as bisector velocities, or measured velocities of H$\alpha$ rather than
metallic lines \citep{1992MNRAS.259..474W}. The fact that the updated result
presented here is consistent with all other data spanning nearly 40 years,
including Petterson et al.'s post-1996 data strongly supports this conjecture. 
In addition, we point out the order of magnitude smaller residual scatter in
Fig.\,\ref{fig:YZCar_Orbit} compared to the scatter of residuals shown in
\citet[Fig.\,5]{2004MNRAS.350...95P}.

Using this updated orbital solution together with the actual dates of
the {\it HST} spatial scan observations, we determine $\hat{\varpi}\sin{i} = 0.26$\,AU. Assuming an average inclination of $60$\,degrees, this would lead to a parallax error of up to $30\%$ (e.g., $\pm 100\,\mu$arcsec at $3$\,kpc distance), which should be clearly
noticeable in the astrometric measurements.  We therefore hope to
obtain $2$ additional epochs of spatial scan observations of
YZ\,Carinae in order to improve the parallax measurement by accounting
for orbital motion in the astrometric model.

\floattable
\begin{deluxetable}{lrrr}
\tablecaption{Orbital elements for YZ\,Carinae}
\tablehead{ \colhead{Parameter} & \colhead{C83$^1$} & \colhead{P04$^2$} &
\colhead{This work} }
\startdata
\Porb\ [d] & $850 \pm 11$ & $657.3 \pm 0.3$ & $830.22 \pm 0.34$ \\
$e$ & $0.13 \pm 0.07$ & $0.14 \pm 0.03$ & $0.041 \pm 0.010$ \\
$K$ [\kms] & $9.4\pm0.5$ & $10.0 \pm 0.4$ & $10.26 \pm 0.82$ \\
$v_\gamma$ [\kms] & $1.0\pm0.4$ & $0.0 \pm 0.2$ & $0.844 \pm 0.063$ \\
$T_0$ - 2.4M & $43575 \pm 11$ & $42250 \pm 9$ & $53422\pm 29$ \\
$\omega$ [deg] & $239 \pm 6$ & $116 \pm 5$ & $195 \pm 12$ \\
\asini\ [$10^6$ km] & $-$ & $89$ &  $117.1 \pm 9.4$ \\
$f_{\rm{mass}}$ [\Msol] & 0.071 & $0.066$ & $0.093 \pm 0.041$ \\
$i_{\rm{min}}$ [deg] & & & $22^\dagger$ \\
rms [\kms] & $-$ & $-$ & $0.39$ \\
$\hat{\varpi}\sin{i} \ [\%\ \varpi$] & $-$ & $-$ & $26$ \\
$\hat{\varpi}\sin{i} / \sin{i_{\rm{min}}}\ [\%\ \varpi$] & & & 76 \\
\enddata
\tablereferences{$^1$: \citet{1983MNRAS.205.1135C}, $^2$:
\citet{2004MNRAS.350...95P}.}
\tablecomments{Some uncertainties derived here are larger than in the
literature due to the combined pulsation plus orbital fit. The minimum
inclination (marked by $^\dagger$) is determined from the mass function assuming
$m_1 = 7\,M_\odot$.
For $m_1=9\,M_\odot$, $\min{i}=20$\,deg.}
\label{tab:YZCarOrbit}
\end{deluxetable}

\subsection{Long-timescale variations of $v_\gamma$}
\label{sec:LongTimeScaleOrbits}
Whereas the above sections focus on relatively short timescale orbital motion
(\Porb$\lesssim 5\,$yr), we also investigate longer-term variations of
$v_\gamma$, which are usually interpreted as evidence for spectroscopic
binarity, by comparing our data with older measurements from the literature.

To this end, we divide the available data\hbox{---}literature RVs and our new
RVs\hbox{---}into tranches that provide adequate phase sampling, balancing
better phase coverage against temporal baseline per tranche. We then determine
best-fit \Ppuls, epoch of minimum RV, and $v_\gamma$ for the data
corresponding to each tranche using an RV template fitting approach.
To achieve an accurate result it is crucial for the data belonging to a given
tranche to sample both the rising and falling branch of the RV curve. The shaded
regions in Fig.\,\ref{fig:OMCvsBJD} indicate how data tranches were selected for
our new data; literature data were done analogously by inspection of the
available data.

The RV templates used to fit each data tranche are created as Fourier series
with harmonic coefficients resulting from the pulsational RV curve modeling
described in \S\ref{sec:RVvariability}. Each template fit determines two
quantities for a fixed pulsation period \Ppuls$(t)$: $v_\gamma (t)$, and a phase
offset $\delta \phi(t)$ relative to the mean observation date required to
determine the time of minimum RV ($E(t)$) corresponding to this tranche and
\Ppuls$(t)$. To account for period changes, we determine the globally
best-fitting (minimum $\chi^2$) solution for a grid of input \Ppuls\ that lie
within $0.1$\,d of the value listed in Tab.\,\ref{tab:RVresults}. We then repeat
this procedure to within $0.01$\,d around the previous best-fit period to
achieve a finer result. The final result of each fitted tranche is visually
inspected to ensure a satisfactory result.

The main limitations of using time-variable $v_\gamma$ as an indicator of
spectroscopic binarity are 1) RV zero-point offsets among spectrographs and
authors \citep[up to several hundred \ms, see][]{2015AJ....150...13E}; 2)
non-linear period fluctuations preventing adequate phase-folding of a given tranche's data
(can be on the order of $1$\,\kms, cf. AQ\,Pup in \S\,\ref{sec:oldbinaries}); 3)
apparent changes in in $v_\gamma$ induced by cycle-to-cycle changes of RV
variability \citep[up to a few hundred \ms, see][]{2016MNRAS.455.4231A}.
Determining the impact of 1) would require precision standard star RV time-series that are
generally not available in the literature. Points 2) and 3) are particularly
relevant for long-period (\Ppuls$ \gtrsim 20$\,d) Cepheids as explained in
\S\ref{sec:caveats}. To avoid spurious detections, we therefore consider the
overall behavior of $v_\gamma(t)$ over all tranches and adopt a threshold of
$1\,$\kms\ as the minimum offset before concluding on spectroscopic binarity.

\begin{figure*}
\caption{Pulsation-average velocities $v_\gamma (t)$ determined by fitting
newly-created RV curve templates.
XZ\,Car, AQ\,Car, CD\,Cyg are newly-discovered
spectroscopic binary Cepheids, see \S\ref{sec:newbinaries}. SS\,CMa, VY\,Car,
KN\,Cen, AQ\,Pup, SZ\,Cyg, X\,Pup, and WZ\,Sgr have been reported as such in the
literature and are discussed in \S\ref{sec:oldbinaries}. The red dashed line
indicates 0, whereas the green dotted line shows an offset of $1\,$\kms,
usually taken as indicative of a variation in $v_\gamma$ due to binarity, cf.
\S\ref{sec:LongTimeScaleOrbits}.}
\centering
\includegraphics{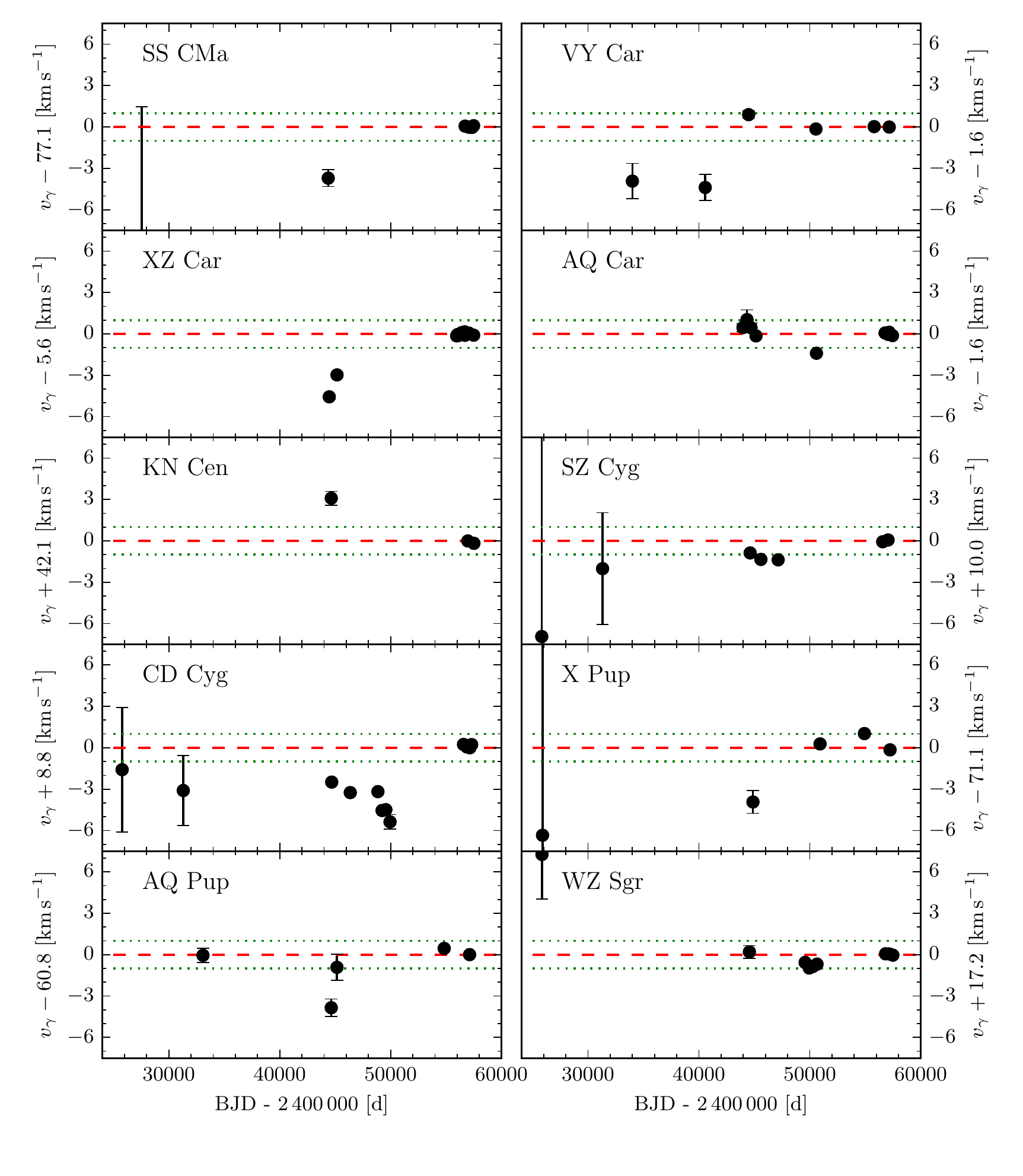} 
\label{fig:deltavgamma}
\end{figure*}

\begin{figure*}
\caption{Pulsation-average velocities $v_\gamma (t)$ of Cepheids not exhibiting
significant variations in $v_\gamma$ that have not previously been reported to
be spectroscopic binaries. HW\,Car is not shown here due to a lack of literature
data. For VX\,Per, the epoch near JD 2\,443\,000 illustrates the range of
possible RV zero-point offsets among instruments via the difference in
$v_\gamma$ inferred using contemporaneous data by \citet{1999A+AS..140...79I}
and \citet{1988ApJS...66...43B}, the latter of which yield a value lower by $\sim
0.8$\,\kms.} 
\centering
\includegraphics{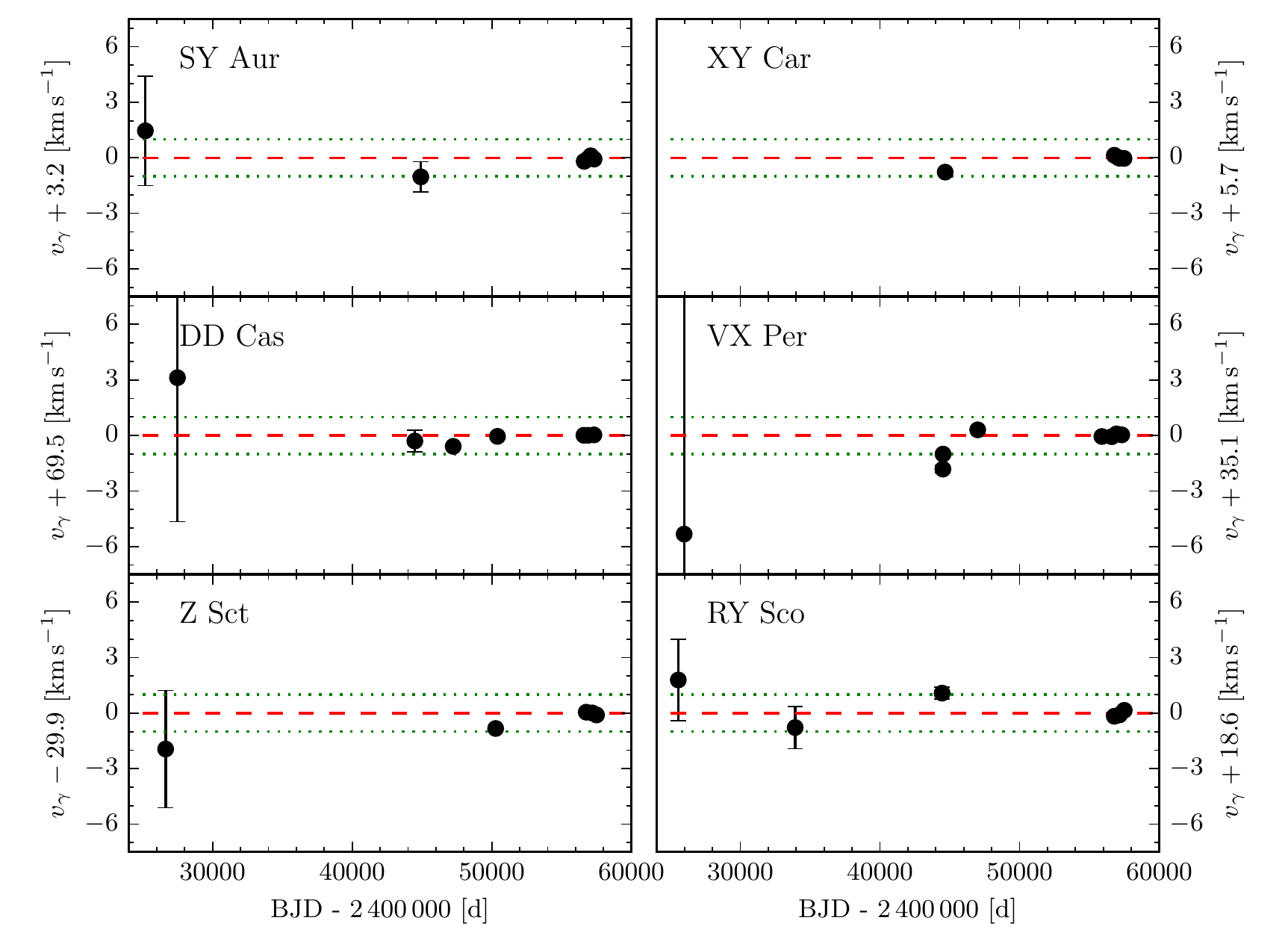} 
\label{fig:deltavgamma_stable}
\end{figure*}

We have thus investigated possible long-term variations of $v_\gamma$ for 15 of
our 18 Cepheids, excluding YZ\,Car (\S\ref{sec:YZCar}), HW\,Car and S\,Vul
(both: lack of literature data).
In the following, we report the discovery of three new candidate spectroscopic
binaries (\S\ref{sec:newbinaries}), followed by a critical investigation of
Cepheids previously reported to be spectroscopic binaries
(\S\ref{sec:oldbinaries}).
Cepheids reported here or in the literature to be binaries are shown in
Fig.\,\ref{fig:deltavgamma}, whereas Fig.\,\ref{fig:deltavgamma_stable} presents 
Cepheids for which no significant variations in $v_\gamma$ were
found and that have not previously been reported to be binaries.

\subsubsection{New candidate spectroscopic binaries}
\label{sec:newbinaries}

\floattable
\begin{deluxetable}{lrrrrrrrr}
\tablecaption{Time-variable $v_\gamma$ of new spectroscopic binaries based on
RV template fitting}
\tablehead{ \colhead{Cepheid} & \colhead{\Ppuls} & \colhead{Epoch} &
\colhead{$\Delta t$} & \colhead{$N_{\rm{RV}}$} & 
\colhead{$v_\gamma$} & \colhead{stdmer} &
\colhead{$\sigma(v_\gamma)$} &
\colhead{References}
\\
& \colhead{[d]} & \colhead{BJD$-2.4$M} & \colhead{[yr]} & & \colhead{[\kms]} &
\colhead{[\kms]} & \colhead{[\kms]} }
\startdata
XZ Car  &  16.6525  &  44464.9312  &  2.21  &  38  &  1.07  &  0.22  & 
0.16 &  CCG \\
XZ Car  &  16.6565  &  45164.0712  &  0.98  &  13  &  2.66  &  0.41  & 
0.18 &  CCG; PBM \\
XZ Car  &  16.6573  &  55955.0183  &  0.05  &  15  &  5.49  &  0.36  & 
0.003  &  h \\
XZ Car  &  16.6507  &  56021.6499  &  0.04  &  9  &  5.58  &  0.23  & 
0.003  &  h \\
XZ Car  &  16.6495  &  56054.9312  &  0.04  &  11  &  5.55  &  0.26  & 
0.003  &  h \\
XZ Car  &  16.6549  &  56088.2126  &  0.04  &  5  &  5.51  &  0.27  & 
0.001  &  h \\
XZ Car  &  16.6523  &  56404.6109  &  0.33  &  12  &  5.72  &  0.18  & 
0.001  &  h \\
XZ Car  &  16.6473  &  56671.0475  &  0.04  &  12  &  5.78  &  0.36  & 
0.002  &  h \\
XZ Car  &  16.6473  &  56704.3392  &  0.04  &  14  &  5.53  &  0.39  & 
0.004  &  h \\
XZ Car  &  16.6569  &  56787.6148  &  0.04  &  14  &  5.63  &  0.26  & 
0.001 &  h \\
XZ Car  &  16.6549  &  57054.0691  &  0.04  &  19  &  5.69  &  0.21  & 
$<0.001$  &  h \\
XZ Car  &  16.6541  &  57486.9476  &  0.03  &  7  &  5.54  &  0.46  &  0.011
& h \\
\hline
AQ Car  &  9.7745  &  43975.7970  &  0.02  &  8  &  2.03  &  0.32  & 
0.27 &  CCG \\
AQ Car  &  9.7717  &  44337.1307  &  0.58  &  18  &  2.61  &  0.49  & 
0.70  &  CCG \\
AQ Car  &  9.7689  &  44698.7272  &  0.51  &  25  &  2.02  &  0.32  & 
0.32  &  CCG \\
AQ Car  &  9.7679  &  45148.1286  &  1.32  &  19  &  1.43  &  0.17  & 
0.018 &  CCG; PBM \\
AQ Car  &  9.7677  &  50599.3879  &  0.83  &  15  &  0.17  &  0.21  & 
0.015 &  B02 \\
AQ Car  &  9.7679  &  56793.2563  &  0.03  &  13  &  1.64  &  0.20  & 
0.001 & h \\
AQ Car  &  9.7705  &  57057.0354  &  0.04  &  21  &  1.54  &  0.14  & 
$<0.001$  &  h \\
AQ Car  &  9.7673  &  57154.7358  &  0.03  &  13  &  1.69  &  0.15  & 
0.001  &  h \\
AQ Car  &  9.7698  &  57457.5769  &  0.11  &  12  &  1.45  &  0.26  & 
0.002  &  h \\
\hline
CD Cyg  &  17.075  &  25787.5216  &  5.19  &  12  &  -10.39  &  1.07  & 
4.51 &  J37 \\
CD Cyg  &  17.081  &  31301.5552  &  0.07  &  22  &  -11.90  &  0.63  & 
2.54  &  S45 \\
CD Cyg  &  17.0759  &  44686.6171  &  2.66  &  30  &  -11.29  &  0.22  & 
0.020  &  I99; B88 \\
CD Cyg  &  17.0758  &  46359.9408  &  3.54  &  10  &  -12.05  &  0.38  & 
0.044  &  I99 \\
CD Cyg  &  17.079  &  48852.8005  &  0.27  &  19  &  -11.98  &  0.37  & 
0.051  &  G92 \\
CD Cyg  &  17.071  &  49228.4290  &  0.15  &  6  &  -13.35  &  0.66  & 
0.093  &  G92 \\
CD Cyg  &  17.071  &  49569.8434  &  0.22  &  9  &  -13.29  &  0.36  & 
0.055  &  G92 \\
CD Cyg  &  17.081  &  49945.3098  &  0.12  &  12  &  -14.17  &  0.58  & 
0.52  &  G92 \\
CD Cyg  &  17.0792  &  56570.5614  &  0.36  &  30  &  -8.56  &  0.09  & 
0.001  &  h \\
CD Cyg  &  17.0748  &  56877.9332  &  0.34  &  39  &  -8.74  &  0.13  & 
0.001  &  h \\
CD Cyg  &  17.0742  &  57151.1361  &  0.19  &  17  &  -8.80  &  0.19  & 
0.003 &  h \\
CD Cyg  &  17.0752  &  57304.8482  &  0.25  &  25  &  -8.57  &  0.13  & 
0.001 &  h \\
\hline
\enddata
\tablereferences{CCG: \citet{1985ApJS...57..595C}, 
PBM: \citet{1994A+AS..105..165P}, B02: \citet{2002ApJS..140..465B}, 
J37: \citet{1937ApJ....86..363J}, I99: \citet{1999A+AS..140...79I}, 
B88: \citet{1988ApJS...66...43B}, 
S45: \citet{1945ApJ...102..232S}, 
G92: \citet{1992SvAL...18..316G}, 
h: this work.}
\tablecomments{\Ppuls, E, and $v_\gamma$ are based on RV template fitting.
$\Delta t$ indicates the timespan of the measurement, $N_{\rm{RV}}$ the number
of measurements fitted, `stdmer' the standard mean error based on residual
scatter, $\sigma(v_\gamma)$ the uncertainty from the fit covariance matrix.}
\label{tab:newbinaries}
\end{deluxetable}

Based on our RV template fitting approach, we report the discovery of three new
spectroscopic binary candidates: XZ\,Car, AQ\,Car, and CD\,Cyg; 
Table\,\ref{tab:newbinaries} lists these results.

Column `stdmer' quotes the
standard mean error based on the residual scatter of the fit and should be
compared to the fit uncertainty derived from the covariance matrix. It is
interesting to note that `stdmer' tends to be smaller than $\sigma(v_\gamma)$
for older, imprecise data, whereas the opposite is the case for new high-precision
data. Specifically, the improvement of `stdmer' stagnates compared to the
improvement in $\sigma(v_\gamma)$ when using more precise (newer) data to
determine $v_\gamma$. This is a consequence of the intrinsic astrophysical noise
of Cepheid pulsations that manifest as fluctuations in period and RV curve shape
\citep{2014A+A...566L..10A,2016MNRAS.455.4231A}.
Inspection of the $v_\gamma$ values derived for XZ\,Car shows that this
astrophysical noise can lead to variations larger than a factor of several
$\sigma(v_\gamma)$. For XZ\,Car specifically, the (unweighted) mean $v_\gamma$
inferred by template fitting of exclusively new measurements is $v_\gamma =
5.603 \pm 0.030$\,\kms, which is very close to the value of $v_\gamma$
determined in a combined Fourier fit (5.671\,\kms, cf.
Tab.\,\ref{tab:RVresults}).

Adopting the above-stated threshold of $1$\,\kms, we find that XZ\,Carinae,
AQ\,Carinae, and CD\,Cygni exhibit significantly time-dependent $v_\gamma$ on
time scales longer than a few years. This marks the first discovery of XZ\,Car's
binarity, and our new data are decisive in demonstrating the likely binary
nature of CD\,Cyg, which was previously considered not to be a spectroscopic
binary \citep{2015AJ....150...13E}. While AQ\,Car's comparatively small
$v_\gamma$ variation ($\sim 1.5\,$\kms) among 
\citep{2002ApJS..140..465B} and our RVs renders this evidence
tentative, we note that the larger difference to older RVs
\citep[$> 2$\,\kms][]{1985ApJS...57..595C,1994A+AS..105..165P} and in particular
the high quality of the \citet{2002ApJS..140..465B} data that provide a
well-constrained fit corroborate an interpretation as evidence for
spectroscopic binarity.

We note that none of these three Cepheids are expected to have incurred
significant parallax error due to orbital motion, Tab.\,\ref{tab:upperlimits}
listing $\leq 1.6\%$ for each of their respective
$\rm{max}(\hat{\varpi}\sin{i})$ solutions.
However, proper motions estimated using long temporal baselines such as the Tycho-{\it Gaia} astrometric solution \citep{2015A+A...574A.115M} may be affected by 
long-timescale orbital motion.

\subsubsection{Revisiting previously reported spectroscopic binary candidates}
\label{sec:oldbinaries}

In addition to the new binaries presented above, Figure\,\ref{fig:deltavgamma}
also shows $v_\gamma(t)$ of seven Cepheids that have previously been considered
to be spectroscopic binaries.

We have recently discussed the binarity of SS\,Canis\,Majoris in light of our
{\it HST} astrometric and recent high-precision RV measurements
\citep{2016ApJ...825...11C}. Using our RV template fitting technique, we here
additionally investigate long-term variations of $v_\gamma$ not discussed in our
preceding paper. We find that the oldest data by \citet{1937ApJ....86..363J} are
not sufficiently accurate to determine $v_\gamma$ with precision, although the
central values of our fit results do reproduce the difference of $\sim
15\,$\kms\ compared to RV data by \citet{1985SAAOC...9....5C} as previously
reported \citep{1996A+A...311..189S}.
Using Coulson \& Caldwell's data, we determine a significant offset in
$v_\gamma$ of $\sim 3.8\,$\kms\ compared to our new data, cf.
Tab.\,\ref{tab:oldbinaries}, which would support a long-timescale spectroscopic
binary nature of SS\,CMa.

KN\,Centauri is a special case among binary Cepheids in that its hot main
sequence companion has been detected and characterized using optical photometry
\citep{1977MNRAS.178..505M,1980PASP...92..315M} as well as UV
\citep{1985ApJ...296..175B,1994ApJ...436..273E} and optical
\citep{1980SAAOC...1..257L} spectra. However, the orbital signature on
the RV curve had thus far escaped detection. We here report a detection of this
signature based on the $\sim 3\,$\kms offset in $v_\gamma$, cf.
Tab.\,\ref{tab:oldbinaries}. The potential ability to detect the orbital motion
of both components separately renders KN\,Cen an important target for a future
model-independent mass measurement. 

VY\,Carinae is a cluster Cepheid \citep{1977AJ.....82..163T,2013MNRAS.434.2238A}
whose previously reported spectroscopic binarity was based on evidence for
low-amplitude variations of $v_\gamma$ \citet[$2K <
5$\,\kms]{1996A+A...311..189S}. The variation of $v_\gamma$ for VY\,Car shown in
Fig.\,\ref{fig:deltavgamma} is peculiar: the two oldest epochs indicate a rather
constant velocity, and so the four newer epochs. At present, it is difficult to
judge whether this pattern is caused by orbital motion or rather due to
systematics such as data quality or sampling.

SZ\,Cygni had previously been reported to exhibit time-variable
$v_\gamma$ by \citet{1945ApJ...102..232S} and \citet{1991CoKon..96..123S}.
However, we find that the data upon which this evidence was based do not
constrain $v_\gamma$ at the crucial epochs. Notably, the data by
\citet{1937ApJ....86..363J} and \citet{1945ApJ...102..232S} are imprecise and do
not sample pulsation phase very well. We do, however, find that data by
\cite{1999A+AS..140...79I} indicate a small difference of about $1.3\,$\kms\
relative to $v_\gamma$ determined using our new measurements.

X\,Puppis was first reported to exhibit time-variable $v_\gamma$
by \citet{2012MNRAS.426.3154S}. Similarly to SZ\,Cygni, we find that the oldest
RV data are insufficiently accurate and sampled to constrain the fit well.
However, data published by \citet{1988ApJS...66...43B} and
\citet{2001JAD.....7....4C} yield a significantly lower $v_\gamma$ than newer
measurements, which furthermore exhibit a suspicious trend in $v_\gamma(t)$.
Additional high-precision observations of X Pup taken over the next few years
will clarify whether this variation is consistent with orbital motion.

\floattable
\begin{deluxetable}{lrrrrrrrr}
\tablecaption{Time-dependent $v_\gamma$ based on RV template fitting for
reported binary Cepheids}
\tablehead{ \colhead{Cepheid} & \colhead{\Ppuls} & \colhead{Epoch} &
\colhead{$\Delta t$} & \colhead{$N_{\rm{RV}}$} & 
\colhead{$v_\gamma$} & \colhead{stdmer} &
\colhead{$\sigma(v_\gamma)$} &
\colhead{References} \\
& \colhead{[d]} & \colhead{BJD$-2.4$M} & \colhead{[yr]} & & \colhead{[\kms]} & \colhead{[\kms]} & \colhead{[\kms]} & }
\startdata
SS CMa  &  12.3568  &  27554.7032  &  3.15  &  5  &  58.36  &  2.51  & 
20.2  &  J37 \\
SS CMa  &  12.3536  &  44373.6734  &  2.04  &  47  &  73.39  &  0.45  &  0.61 &  CC85 \\
SS CMa  &  12.3528  &  56716.3689  &  1.58  &  27  &  77.16  &  0.10  &  0.002 &  h \\
SS CMa  &  12.3522  &  57062.2692  &  0.12  &  26  &  77.08  &  0.17  & 
0.001  &  h \\
SS CMa  &  12.3478  &  57346.3907  &  0.06  &  19  &  77.07  &  0.21  &  0.002  &  h \\
SS CMa  &  12.3478  &  57482.2462  &  0.02  &  8  &  77.19  &  0.49  &  0.014  &  h \\
\hline
VY Car  &  18.8865  &  33999.9783  &  1.04  &  15  &  -2.31  &  0.59  & 
1.28  &  S55 \\
VY Car  &  18.8847  &  40567.9096  &  1.08  &  6  &  -2.77  &  0.64  & 
0.95 &  L80 \\
VY Car  &  18.8853  &  44483.7134  &  3.08  &  60  &  2.51  &  0.30  & 
0.27  &  CC85 \\
VY Car  &  18.8841  &  50551.7748  &  0.82  &  16  &  1.47  &  0.45  & 
0.064  &  B02 \\
VY Car  &  18.8831  &  55803.4453  &  1.21  &  41  &  1.64  &  0.41  & 
0.002  &  h \\
VY Car  &  18.8825  &  57162.9905  &  1.19  &  42  &  1.61  &  0.26  & 
0.001 &  h \\
\hline
KN Cen  &  34.0232  &  44545.6646  &  3.08  &  34  &  -39.05  &  0.41  & 
0.50  &  CC85 \\
KN Cen  &  34.0192  &  56965.1895  &  1.05  &  47  &  -42.15  &  0.48  &  
0.008  & h \\
KN Cen  &  34.0206  &  57509.4973  &  0.18  &  23  &  -42.32  &  0.98  &  
0.036  & h \\
\hline
SZ Cyg  &  15.1127  &  25826.3093  &  7.45  &  8  &  -16.97  &  2.26  & 
19.3 &  J37 \\
SZ Cyg  &  15.1063  &  31310.5534  &  0.07  &  17  &  -12.05  &  0.74  & 
4.06  &  S45 \\
SZ Cyg  &  15.1079  &  44622.2780  &  2.48  &  28  &  -10.91  &  0.26  & 
0.038 &  I99;  B88 \\
SZ Cyg  &  15.1112  &  45589.2646  &  3.02  &  18  &  -11.38  &  0.17  & 
0.012 &  I99 \\
SZ Cyg  &  15.1113  &  47145.6313  &  3.29  &  14  &  -11.41  &  0.22  & 
0.016  & I99 \\
SZ Cyg  &  15.1157  &  56574.1255  &  0.36  &  30  &  -10.10  &  0.11  & 
0.002  &  h \\
SZ Cyg  &  15.1113  &  57057.7047  &  1.36  &  77  &  -9.97  &  0.05  & 
$<0.001$  & h \\
\hline
X Pup  &  25.9582  &  25896.3080 &  3.98  &  10  &  64.78  &  1.54  & 
14.6 &  J37 \\
X Pup  &  25.9628  &  44874.5896  &  2.33  &  32  &  67.19  &  0.49  & 
0.82  &  B88; C01 \\
X Pup  &  25.9608  &  50924.8055  &  2.10  &  33  &  71.40  &  0.33  & 
0.034 &  B02; P05 \\
X Pup  &  25.9605  &  54924.3983  &  1.23  &  42  &  72.15  &  0.37  & 
0.017 &  S11 \\
X Pup  &  25.9602  &  57235.7830 &  1.38  &  84  &  70.97  &  0.25  & 
0.002 &  h \\
\hline
AQ Pup  &  30.1768  &  33078.6817  &  0.45  &  13  &  60.75  &  0.37  & 
0.52  &  S55 \\
AQ Pup  &  30.1768  &  44647.4321  &  0.49  &  30  &  56.95  &  0.43  &  0.63 &  CC85; B88 \\
AQ Pup  &  30.1834  &  45159.8310  &  1.21  &  14  &  59.88  &  0.49  &  0.94 &   CC85; B88 \\
AQ Pup  &  30.1806  &  54829.5366  &  1.74  &  38  &  61.25  &  0.70  & 
0.068 &  S11 \\
AQ Pup  &  30.1820  &  57093.2976  &  2.96  &  98  &  60.80  &  0.34  & 
0.006 &  h \\
\hline
WZ Sgr  &  21.8521  &  25849.9613  &  5.09  &  9  &  -9.91  &  0.92  & 
3.23  &  J37 \\
WZ Sgr  &  21.8549  &  44553.0330  &  2.83  &  24  &  -16.97  &  0.37  & 
0.46 &  CC85; B88 \\
WZ Sgr  &  21.8461  &  49578.7275  &  0.17  &  26  &  -17.74  &  0.30  & 
0.12  &  G92 \\
WZ Sgr  &  21.8561  &  49950.1832  &  0.27  &  19  &  -18.14  &  0.39  & 
0.05 &  G92 \\
WZ Sgr  &  21.8463  &  50277.9982  &  0.22  &  38  &  -18.03  &  0.25  & 
0.02 &  B02; G92 \\
WZ Sgr  &  21.8535  &  50649.5795  &  0.49  &  32  &  -17.86  &  0.25  & 
0.02  &  B02; G92 \\
WZ Sgr  &  21.8521  &  56833.7761  &  0.23  &  33  &  -17.11  &  0.22  & 
0.001  &  h \\
WZ Sgr  &  21.8507  &  57161.5552  &  0.18  &  29  &  -17.12  &  0.36  & 
0.006  &  h\\
WZ Sgr  &  21.8516  &  57489.2777  &  0.18  &  13  &  -17.21  &  0.30  & 
0.005  &  h \\
\enddata
\tablereferences{S45: \citet{1945ApJ...102..232S}, 
CC85: \citet{1985SAAOC...9....5C},
L80: \citet{1980SAAOC...1..257L},
B88: \citet{1988ApJS...66...43B},
I99: \citet{1999A+AS..140...79I},
S11: \citet{2011A+A...534A..94S},
G92: \citet{1992SvAL...18..316G},
B02: \citet{2002ApJS..140..465B},
J37: \citet{1937ApJ....86..363J},
S55: \citet{1955MNRAS.115..363S},
P05: \citet{2005MNRAS.362.1167P},
C01: \citet{2001JAD.....7....4C} }
\tablecomments{See also Tab.\,\ref{tab:newbinaries}. J37 data do not constrain the fit well for SS\,CMa, SZ\,Cyg, and X\,Pup.}
\label{tab:oldbinaries}
\end{deluxetable}

AQ\,Puppis exhibits very significant non-linear changes of \Ppuls\
\citep{1991Ap+SS.183...17V} in addition to exceptionally fast
($300\,\rm{s\,yr^{-1}}$) secular changes of \Ppuls\ \citep{2000ASPC..203..244B}
that approach values predicted for Cepheids on a first crossing of the instability
strip \citep{2012AJ....144..187T,2016A+A...591A...8A}.
\citet{1966AJ.....71..999F} reported a chance alignment with an OB association
\citep[see also][]{2012AJ....144..187T}, which the {\it HST} spatial scans will
further illuminate.
\citet{1980PASP...92..315M} presented evidence of a photometric companion based
on both a phase-shift and color-loop methodology. AQ\,Pup's spectroscopic binary
nature was reported by \citep{1991Ap+SS.183...17V} based on a systematic offset
in mean velocity between the data from \citet{1937ApJ....86..363J} and those by
\citet{1955MNRAS.115..363S}, \citet{1988ApJS...66...43B}, and
\citet{1985SAAOC...9....5C}.
Ignoring the imprecise and extremely sparse (5 measurements over 4 years) data
by \citet{1937ApJ....86..363J}, we find that nearly all available RV data is
consistent with a constant $v_\gamma$. However, RVs measured near epoch JD
$44650$ appear to be offset by $\sim 3\,$\kms\ (cf.
Tab.\,\ref{tab:oldbinaries} and Fig.\,\ref{fig:deltavgamma}) from measurements
taken just one year later by the same authors
\citep{1985SAAOC...9....5C,1988ApJS...66...43B}. Following visual inspection
of the RV data and given that our new RV data do not indicate fast and
significant variations in $v_\gamma$, we do not consider this apparent offset in
$v_\gamma$ to be a solid indication of spectroscopic binarity. Rather, it
appears more likely that non-linear changes of \Ppuls\ \citep[see
e.g.][Fig.\,10]{2012AJ....144..187T} lead to problems in phase-folding the data,
which is required to determine $v_\gamma$.

WZ\,Sagittarii is a member of the open cluster Turner\,2 
\citep{1993ApJS...85..119T,2013MNRAS.434.2238A}, whose other members may aid in
the determination of its accurate parallax. A
spectroscopic binary nature of WZ\,Sgr has both been argued for 
\citep{1989CoKon..94.....S} and against \citep{2015AJ....150...13E}.   
As Fig.\,\ref{fig:deltavgamma} shows, nearly all RV data are nicely consistent
with a stable $v_\gamma$, the oldest data by \citet[8
measurements]{1937ApJ....86..363J} being the exception. We thus conclude that
WZ\,Sgr is unlikely to be a spectroscopic binary.

\section{Discussion}\label{sec:discussion}
The {\it Gaia} mission is currently measuring highly accurate parallaxes for
hundreds of Galactic Cepheids.
This order-of-magnitude increase in sample size compared to the current $12$ accurately known Cepheid parallaxes \citep{2007AJ....133.1810B,2014ApJ...785..161R,2016ApJ...825...11C} will enable future calibrations
of the extragalactic distance scale based on subsamples of Cepheids selected according to properties deemed particularly suitable for this task.
To this end, a detailed vetting process that considers the wide range of
information available for Galactic Cepheids is required. We consider the vetting
process of Galactic  Leavitt law calibrators to be a crucial step towards a measurement of $H_0$ with $1\%$ accuracy. 

The investigation of binarity is an important element of this vetting process. In this work, we focus on the contribution that RV measurements can make in this
regard. Specifically, we use RVs to constrain possible parallax error due to orbital motion for our {\it HST} parallaxes \citep{2014ApJ...785..161R,2016ApJ...825...11C}. This is very important, since the typically 5 observed {\it HST} epochs do not provide sufficient degrees of freedom to determine position, proper motion, parallax\footnote{note that five degrees of freedom are sufficient to constrain these parameters, since spatial scan measurements are one-dimensional by construction, cf. \citet{2014ApJ...785..161R}} \emph{and} orbital motion simultaneously. Hence, this work informs the systematic uncertainty budget of the {\it HST} spatial scan parallaxes and increases confidence in their accuracy. 

Our work demonstrates that RVs are very well-suited for investigating this parallax error, since they provide tight constraints on the range of orbital periods that would most impact the parallax measurements ($1-3$\,yr), despite their insensitivity to inclination. This is because the orbital RV signal for a given \Porb\ and $e$ depends on the mass function ($(m_2 \sin{i})^3/(m_1 + m_2)^2$) of the binary. Another method capable of investigating such short-period systems is optical/NIR long-baseline interferometry. However, the ability to detect companion stars interferometrically depends on the luminosity contrast \citep[feasible dynamic range of $1:200$]{2015A+A...579A..68G}, which tends to be very high due to the evolutionary differences between a Cepheid and its typically main sequence companion, cf. \S\ref{sec:photometric}, as well as the nature of the mass-luminosity relation.

Our RV-based results presented here indicate that orbital motion-induced parallax error is insignificant for most (18 of the total 19) Cepheids in the sample thanks to tight upper limits on undetected orbital configurations. 
Since it is highly unlikely for a large fraction of Cepheids to have nearly face-on orbits ($1/\sin{i} < 3$ for 94\,\% of possible $i$ values), we do not expect more than one of these 18 Cepheids to be subject to (projection-corrected) parallax error due to orbital motion exceeding $\pm 10$\,\% and we have at present no indication of any such error.

The exception among our sample stars is YZ\,Carinae whose orbit is clearly detected and expected to significantly affect parallax ($\hat{\varpi}\sin{i} \sim \pm 100\,\mu$arcsec) \deleted{bias}. Additional spatial scans will be obtained for this star in order to allow our astrometric modeling to account for orbital motion. Correcting the orbital solution (cf. \S\ref{sec:YZCar}) was crucial to provide adequate constraints to this effect.

The method described here can also be applied to RVs measured using {\it Gaia}'s {\it RVS} instrument provided that time-series RVs are sufficiently precise to provide stringent constraints on undetected orbital configurations. However, {\it Gaia} has the advantage of gathering an average of $\sim 70$ astrometric measurements per star and is therefore able to directly include orbital motion in the astrometric modeling. Further ground-based observations are being secured to assist an investigation of binarity in support of {\it Gaia}.

The long-timescale (\Porb$\gg 5$\,yr) spectroscopic binaries reported in \S\ref{sec:LongTimeScaleOrbits} are not expected to affect our {\it HST} parallax measurements. However, proper motion measurements may be affected by such long-period orbital motion.

Of course, the impact of binarity on the distance scale is
not limited to parallax measurements and stands to benefit from an
investigation of data other than purely RV observations. Conversely, other known
properties of Cepheids (besides binarity) deserve detailed investigation in
terms of their impact on the calibration of the distance scale. In the
following, we provide an overview of considerations to be made by such a vetting
process directly related to the present work.

\subsection{Binary frequency in this sample}\label{sec:disc:binaryfrequency}

The Cepheid binary fraction has been a topic of intense research for several
decades
\citep[e.g.][]{1968MNRAS.141..109L,1977MNRAS.178..505M,1985ApJ...296..175B,1991CoKon..96..123S,1994AJ....108..653E,2015AJ....150...13E,2016AJ....151..108E,2016AJ....151..129E}.
Yet, inspection of the available literature data of previously reported
candidate binaries (cf.
\S\ref{sec:LongTimeScaleOrbits}) and the discovery of previously unknown binary
systems among our relatively bright Cepheids suggests that far from everything
is known for even relatively well-studied cases.

As mentioned in \S\ref{sec:selection}, the present sample of Cepheids is not
random with respect to binarity and may therefore not be representative of the
binary fraction of all Cepheids. Nevertheless, we summarize our investigation of
binarity for the program Cepheids. Convincing evidence for spectroscopic binarity for 5
of 19 Cepheids (SS\,CMa, YZ\,Car, XZ\,Car, KN\,Cen, and CD\,Cyg).
Four additional Cepheids (VY\,Car, AQ\,Car, SZ\,Cyg, X\,Pup)
exhibit tentative evidence of variations in $v_\gamma$ consistent with binarity,
although imprecise literature data and sometimes heavily fluctuating pulsational
variability renders these results inconclusive.
The literature further indicates DD\,Cas to have an unresolved photometric
companions (this applies also to KN\,Cen), bringing the total number of binaries
in this sample to between 6/19 ($32\%$) and 10/19 ($53\%$), depending on the
inclusion of questionable candidates. This is broadly consistent with other
recent estimates, e.g. by \citet[$35\%$]{2015A+A...574A...2N} and \citet[$29\pm
8\%$ for \Porb$<20$\,yr]{2015AJ....150...13E}. These numbers do not include
previously reported cases of visual binaries with extreme separations (greater
than a couple arcseconds) or Cepheids belonging to open clusters.

Furthermore, we stress that not all binary Cepheids are fundamentally unsuitable
as high-accuracy Leavitt law calibrators, provided their photometry is not
biased (cf. \S\ref{sec:photometric}) and that parallax can be measured accurately (cf.
\S\ref{sec:upperlims}). Additional photometric and interferometric observations
will be useful to investigate these points.

\subsection{Photometric bias due to companions}\label{sec:photometric}

The literature contains frequent references to binarity
as being a significant source of photometric bias for the estimation of absolute magnitudes.
For distance scale applications, this leads to two main questions:
1) what is the (pulsation-phase) average contrast between a Cepheid and
a typical companion star? 2) how large of an effect on the distance scale
could result from systematic differences in binary statistics among selected
samples of Galactic and extragalactic Cepheids? 

We therefore estimate the photometric contrast between Cepheids and typical,
spatially unresolved, hot main sequence companions via isochrones computed using
the Geneva stellar evolution group's
\citep{2012A+A...537A.146E,2013A+A...553A..24G} online
tool\footnote{\url{http://obswww.unige.ch/Recherche/evoldb/index/Isochrone/}}.
We estimate the luminosity of the hot companion assuming a fixed mass ratio of
$q = m_2 / m_1 = 0.7$ as a conservative typical value based on the extensive
work by N. Evans and collaborators
\citep[e.g.][]{1995ApJ...445..393E,2013AJ....146...93R}.
For KN\,Cen and DD\,Cas, we use information on detected companions
\citep{1977MNRAS.178..505M} to estimate worst case scenarios.
Specifically, we assume a brighter B2 companion for KN\,Cen, despite {\it IUE}
spectra indicating a B6 dwarf \citep{1994ApJ...436..273E}.
Masses are referred to here as lower case $m$ to distinguish them from
magnitudes $M$. The mass of the Cepheid is determined via a (pulsation)
period-mass relation based on Geneva models \citep{2016A+A...591A...8A} and thus
sets the mass scale for both stars. The age of the isochrone is adopted based on
period-age relations by \citet{2016A+A...591A...8A} using period change
information measured or compiled by \citet{2006PASP..118..410T} to inform the
crossing number, where possible.
The isochrone is computed for the inferred Cepheid's age, solar metallicity, and
typical ZAMS rotation rate ($\omega=0.5$).
The contrast in different pass-bands and filter combinations is  estimated by
forcing the Cepheid to be observed during blue loop evolution and looking up the
properties of the companion with mass close to $m_2$ as per the adopted mass
ratio.
All hypothetical companions thus investigated are hot main sequence stars.

\floattable
\begin{deluxetable}{lccccrrrrrrrrrr}
\tablecaption{Estimating typical photometric bias due to main sequence
companions}
\tablehead{ \colhead{} & \colhead{} & \colhead{} & \colhead{} & \colhead{} &
\multicolumn{6}{c}{---------------- assumes $m_2/m_1 = 0.7$ -------------} \\
\colhead{Cepheid} & \colhead{Xing} & \colhead{age} & \colhead{$\log{\rm{age}}$}
& \colhead{$m_1$} & \colhead{$m_2$} & \colhead{$\Delta M_{\rm{V}}$} & 
\colhead{$\Delta M_{\rm{I}}$} & \colhead{$\Delta M_{\rm{H}}$} &
\colhead{$\Delta W_{\rm{VI}}$} & \colhead{$\Delta W_{\rm{H,VI}}$} } 
\startdata
SY Aur  &  3  &  81  &  7.91  &  6.0  &  4.2  &  -3.89  &  -4.58  &  -5.24  &  -5.65  &  -5.52  \\
SS CMa  &  0  &  68  &  7.83  &  6.3  &  4.4  &  -4.04  &  -4.76  &  -5.44  &  -5.87  &  -5.73  \\
VY Car  &  2  &  52  &  7.71  &  7.3  &  5.1  &  -4.31  &  -5.04  &  -5.75  &  -6.17  &  -6.05  \\
XY Car  &  0  &  67  &  7.83  &  6.3  &  4.4  &  -4.04  &  -4.76  &  -5.44  &  -5.87  &  -5.73  \\
XZ Car  &  0  &  57  &  7.75  &  6.9  &  4.9  &  -4.14  &  -4.87  &  -5.57  &  -6.01  &  -5.87  \\
YZ Car  &  3  &  55  &  7.74  &  6.9  &  4.9  &  -4.14  &  -4.87  &  -5.57  &  -6.01  &  -5.87  \\
AQ Car  &  0  &  78  &  7.89  &  6.0  &  4.2  &  -3.89  &  -4.58  &  -5.24  &  -5.65  &  -5.52  \\
HW Car  &  0  &  80  &  7.91  &  6.0  &  4.2  &  -3.89  &  -4.58  &  -5.24  &  -5.65  &  -5.52  \\
SZ Cyg  &  3  &  62  &  7.8  &  6.6  &  4.6  &  -3.99  &  -4.71  &  -5.4  &  -5.83  &  -5.69  \\
CD Cyg  &  3  &  58  &  7.76  &  6.9  &  4.9  &  -4.14  &  -4.87  &  -5.57  &  -6.01  &  -5.87  \\
VX Per  &  2  &  69  &  7.84  &  6.3  &  4.4  &  -4.04  &  -4.76  &  -5.44  &  -5.87  &  -5.73  \\
X Pup   &  3  &  44  &  7.64  &  7.8  &  5.4  &  -4.42  &  -5.15  &  -5.87  & -6.29  &  -6.17  \\
AQ Pup  &  3  &  40  &  7.6  &  8.2  &  5.7  &  -4.48  &  -5.24  &  -5.98  &  -6.41  &  -6.29  \\
WZ Sgr  &  3  &  49  &  7.69  &  7.3  &  5.1  &  -4.31  &  -5.04  &  -5.75  &  -6.17  &  -6.05  \\
RY Sco  &  3  &  52  &  7.71  &  7.3  &  5.1  &  -4.31  &  -5.04  &  -5.75  &  -6.17  &  -6.05  \\
Z Sct   &  3  &  69  &  7.84  &  6.3  &  4.4  &  -4.04  &  -4.76  &  -5.44  &  -5.87  &  -5.73  \\
S Vul   &  3  &  24  &  7.37  &  10.2  &  7.2  &  -4.55  &  -5.76  &  -7.1  &  -7.63  &  -7.6  \\
\hline
DD Cas (B4V) &  3  &  82  &  7.92  &  6.0  &  4.95  &  -3.21  & 
-3.93  &  -4.62  & -5.05  &  -4.92  \\
KN Cen (B2V) &  3  &  37  &  7.57  &  8.6  &  6.9  &  -3.75  & 
-4.74  &  -5.77  &  -6.28  &  -6.18  \\
\enddata
\tablecomments{Column X marks the instability crossing based on positive
(assumed to be third crossings) and negative (second crossings) observed period changes
\citep{2006PASP..118..410T}. Ages are
estimated using period-age relations for the appropriate crossing assuming
average initial rotation rates \citep{2016A+A...591A...8A}.
Cepheid masses $m_1$ are estimated using isochrones of Solar metallicity 
and average ZAMS rotation \citep{2012A+A...537A.146E,2013A+A...553A..24G} for
a fixed adopted mean color $V-I = 0.5$ during the blue loop phase.
$m_2$ is the mass of a hypothetical companion, where $m_2/m_1 = 0.7$ for most
cases (see \S\ref{sec:photometric}) for the purpose of estimating the contrast
in different filters and filter combinations. DD\,Cas and KN\,Cen are
marked together with companion spectral type estimates \citep{1977MNRAS.178..505M}.
The Cepheids in this program are between $25$ and $80$\,Myr old. Cepheids are much brighter than their main sequence
companions, as expected, and this contrast increases when using longer-wavelength data
and Wesenheit indices, as well as with pulsation period.}
\label{tab:ages}
\end{deluxetable}

Using this approach, we estimate approximate magnitude differences in
bolometric magnitudes, $V$-band, $I$-band, $H$-band, as well as reddening-free
Wesenheit indices
\citep{1982ApJ...253..575M} $W_{VI} = I - 1.55(V-I)$ \citep{2008AcA....58..163S}
and $W_{H,VI} = H - 0.4(V-I)$ \citep{2011ApJ...730..119R}, which are known to be
particularly useful for measuring Cepheid distances using PLRs thanks to reduced
intrinsic PLR dispersion and reduced sensitivity to reddening\footnote{Wesenheit
magnitudes are ``reddening-free'' by construction provided the reddening law is
known sufficiently well}. 

Table\,\ref{tab:ages} lists the results obtained, including the
adopted crossing number as well as inferred age and primary (Cepheid) mass
$m_1$. The remaining columns list the quantities of the companion.

\begin{figure}
\centering
\includegraphics{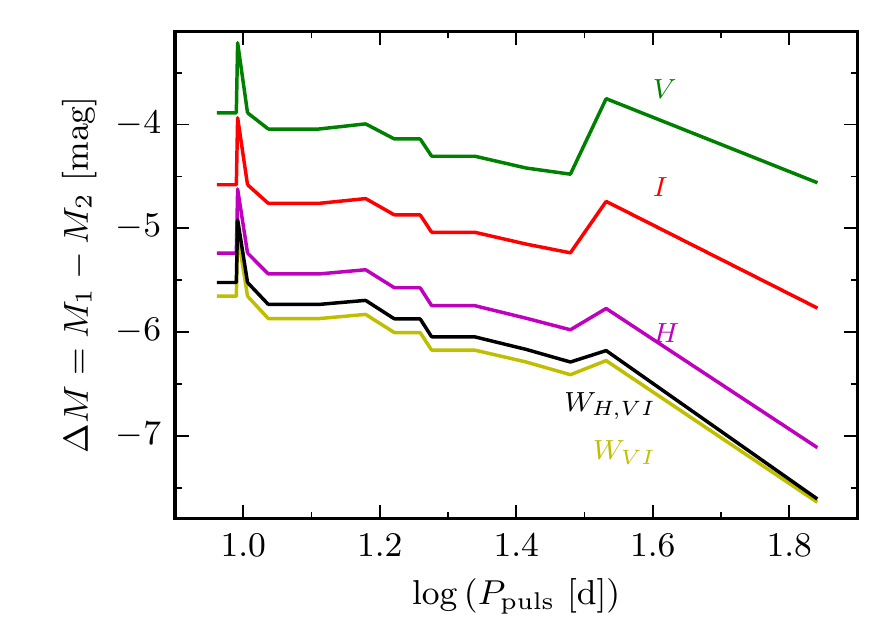}
\caption{Magnitude difference between Cepheid and main sequence companion
(cf. Tab.\,\ref{tab:ages}) in different photometric pass-bands
based on Geneva isochrones.}
\label{fig:photombias}
\end{figure}

Figure\,\ref{fig:photombias} illustrates the comparison. It clearly shows that
the contrast between Cepheid and companion increases with increasing wavelength.
DD\,Cas and KN\,Cen stand out as spikes against the general trend due to higher
mass ratios. Wesenheit indices amplify the contrast since Cepheids tend
to be much redder than their hot main sequence companions.
Fig.\,\ref{fig:photombias} further suggests a \Ppuls-dependence of this
contrast that can be understood via the larger luminosity difference between
stars on main sequence and blue loop evolutionary phases for younger
(higher \Ppuls) stars compared to older (lower \Ppuls) stars.

While a more detailed investigation of this effect would require population
synthesis and examination of Cepheid colors and binary properties, we here note
that the typical contrast using Wesenheit formulations or $H-$band photometry is
on the order of $6$ magnitudes at the typical period ($\log{P_{\rm{puls}}} \sim
1.3$) of extragalactic Cepheid samples (S.~L.~Hoffmann et al., submitted). The
associated increase in brightness of $-0.004$\,mag is much lower than the width
of the instability strip \citep[$0.08$\,mag in $H-$band]{2016arXiv160401424R}
and would lead to a distance error of merely $0.2\,\%$ for an individual
star\hbox{---}much less for an entire population.
Binaries with lower contrast\hbox{---}such as KN\,Cen and DD\,Cas\hbox{---}are
expected to be the exception and even for these cases, no strong photometric
bias is expected in $H-$band. Furthermore, selection criteria applied in the
search for extragalactic Cepheids such as amplitude ratios are expected to
remove very strong outliers \citep[e.g.][]{2015AJ....149..183H}.
While photometric bias is comparatively stronger for shorter-period Cepheids and
in optical pass-bands, we conclude that the light contributed by typical companion
stars is on average negligible for distance scale applications where large
numbers Cepheids are observed using near infrared photometry and/or Wesenheit
magnitudes.

An important additional consideration for the accuracy of
parallax measurements is the phase-dependence of the contrast between a Cepheid
and its spatially unresolved companion that shifts the measured photocenter in
phase with the Cepheid's variability.
This effect has been referred to as ``Variability Induced Movers''
\citep[VIM]{1996A+A...314..679W} in the context of {\it Hipparcos}
\citep{1997ESASP1200.....P,2007ASSL..350.....V}. Whereas most {\it Hipparcos}
VIM solutions were later found to be color-induced \citep{2003A+A...399.1167P},
the correction for VIM-type effects is an integral part of {\it Gaia} data
processing (CU4) due to {\it Gaia}'s much increased astrometric precision. The
same effect can also impact our {\it HST} spatial scanning measurements, depending on i) the
orientation of the scan direction with respect to the orientation of the binary;
ii) the (unresolved) angular separation of the two components; iii) the average
contrast between Cepheid and companion in the passband used to measure parallax;
iv) the pulsation phases sampled by the scan observations. We will study this
effect in detail and constrain its impact on our Cepheid parallaxes in future
work.


\subsection{Peculiar variability}
\label{sec:periodchange}
\label{sec:peculiar}

Cepheids with variability periods longer than $\sim 20$\,d have been shown in
the literature to exhibit non-linear variations of \Ppuls\ and other
cycle-to-cycle modulations
\citep[e.g.][]{2000NewA....4..625B,2014A+A...566L..10A}.
Among the present sample, the most affected stars are KN\,Cen, X\,Pup, AQ\,Pup, and
S\,Vul, with AQ\,Pup having been discussed as a candidate for a first crossing
Cepheid \citep{2016A+A...591A...8A}.

Nonlinear changes of \Ppuls\ complicate the inference of mean
magnitudes from photometric measurements taken at random times using a known pulsation
ephemeris. In the worst case, non-linear period changes may lead to a complete
loss of knowledge of the pulsation phase, leading to observations observed at
random phase. Near-IR photometry can partially mitigate the scatter of the PLR
determined by random-phase observations, since pulsation amplitudes
decrease with increasing wavelength. For a given galaxy, the distance 
error contribution by this term is $< 0.12$\,mag, slightly larger than
the intrinsic dispersion of the Leavitt law in the $H-$band
\citep{2016arXiv160401424R}.

At present, it is not clear what fraction of Cepheids exhibits such effects and
how these phenomena are related to the ability to use affected Cepheids as
precise standard candles. A characterization of non-linear period changes
in extragalactic Cepheids has thus far only been possible in the Magellanic
clouds \citep[S\"uveges \& Anderson,
submitted]{2008AcA....58..313P,2015AcA....65..329S}.
{\it Gaia} parallaxes and the ability to study Galactic Cepheids in great detail will enable
a better understanding of pulsation irregularities and inform the vetting
process of high-accuracy Leavitt law calibrators accordingly.

\section{Conclusions}\label{sec:conclusions}

Over the course of 5 years, we have observed more than $1600$ high-precision RVs
of 19 Galactic classical long-period Cepheids for which the SH0ES team is
measuring highly accurate trigonometric parallaxes using {\it HST/WFC3} spatial
scans \citep{2014ApJ...785..161R,2016ApJ...825...11C}.

We investigate the RV variability of all program Cepheids and construct the most
detailed view of Fourier parameters $R_{21}$, $R_{31}$, $\phi_{21}$, and
$\phi_{31}$ as a function of pulsation period. 

We determine upper limits for undetected spectroscopic companion stars with
\Porb$\lesssim 5$\,yr for 18 of the 19 program Cepheids assuming circular
orbital motion for a range of input values of \Porb. For YZ\,Carinae, we
determine a corrected, full Keplerian orbital solution with \Porb$\sim 830$\,d
(\S\ref{sec:YZCar}).

Using the upper limits on undetected spectroscopic binary configurations in
combination with the properties of the actual {\it HST/WFC3} spatial scan
observations, we compute the absolute inclination-projected maximal parallax error due
to orbital motion, $\hat{\varpi}\sin{i}$, that such ``allowed'' companions could
introduce if the {\it HST} measurements are modeled assuming single star
astrometric models.
We quote the parallax error times $\sin{i}$ to underline that these limits are based
on RV measurements, which cannot constrain inclination.

We exclude significant ($> 2\%$) parallax error due to orbital motion
for the majority of Cepheids with {\it HST} measurements.  We stress
that the values of $\hat{\varpi}\sin{i}$ quoted here are not
indicative of a detected effect on the measured parallax, being entirely limited by the
the available data since no orbital motion was detected for 18 of the
19 Cepheids over the baseline of interest for the parallax
measurements (\Porb$\lesssim 5$\,yr). 

We estimate that YZ\,Carinae's parallax would be affected by approximately $30\%$ ($\sim \pm 100\,\mu$arcsec) if the astrometric measurements were modeled assuming a
single star configuration.
We will therefore obtain additional spatial scan epochs of this star to enable
fitting for the orbit in the astrometric modeling. 

We additionally investigate variations of the pulsation-averaged velocity
$v_\gamma$ to search for indications of possible long-timescale 
(\Porb$\gtrsim 10$\,yr) binarity. We thus report
\begin{itemize}
  \item the discovery of the spectroscopic binary nature of XZ\,Car and CD\,Cyg,
  as well as tentative evidence for AQ\,Car's time-variable $v_\gamma$;
  \item first evidence for orbital motion of KN\,Cen, which has a known
  B-star companion;
  \item a first clear indication of orbital motion for SS\,CMa;
  \item evidence supporting the spectroscopic binarity of VY\,Car and X\,Pup,
  as well as tentative evidence for SZ\,Cyg;
  \item that AQ\,Pup and WZ\,Sgr are likely not to be spectroscopic binaries
  despite previous claims.
\end{itemize}  
Since the associated orbital periods are much longer than the $5$yr
baseline of the {\it HST} spatial scanning observations, no parallax error due to
orbital motion is expected for these stars. The binary fraction in our sample is
$32\hbox{--}52\%$, cf. \S\ref{sec:disc:binaryfrequency}. 

We further discuss the typical photometric impact of unresolved companions based
on stellar isochrones. This leads to the conclusion that near-IR photometry
and/or Wesenheit magnitudes are well-suited to reduce the photometric bias
due to such companions (typical contrast of $\sim 6$\,mag in $H-$band at
$\log{P_{\rm{puls}}} \sim 1.3$).
Moreover, the contrast between a Cepheid and its typical main sequence companion
increases with \Ppuls, i.e., longer-period Cepheids are on average less biased
by flux contributed by unresolved companions. Near-IR photometry is furthermore
well-suited to mitigate PLR scatter in the presence of non-linear fluctuations
of \Ppuls\ thanks to lower IR amplitudes. 

Galactic Cepheids present the unique opportunity to conduct a detailed vetting
of candidates for which the most accurate calibration between pulsation period and
average absolute magnitude can be achieved. In the era of high-accuracy parallax
measurements of long-period Cepheids heralded by {\it Gaia} and the {\it
HST/WFC3} spatial scan observations, such a vetting process may help to
increase the accuracy of the extragalactic distance scale from the bottom up.
Further work along the lines presented here will benefit the overarching
goal of determining the value of the local Hubble constant $H_0$ to $1\%$
accuracy and a better understanding of Dark Energy.

\acknowledgments
We acknowledge observational assistance by Pierre Dubath, Marion Neveu,
Janis Hagelberg, Dominique Naef, Nicolas Cantale, and Malte Tewes. We thank all
Euler and Mercator support staff for their competent assistance. We further thank the referee for a timely and constructive report.

We thank Laszlo Szabados for maintaining the openly accessible database of
Cepheid binaries\footnote{\url{http://www.konkoly.hu/CEP/intro.html}}. This
resource was an invaluable help to track down relevant literature for this
research. We further made use of the DDO Cepheids
database\footnote{\url{http://www.astro.utoronto.ca/DDO/research/Cepheids/}}
\citep{1995IBVS.4148....1F} and the McMaster Cepheid photometry and RV data
archive\footnote{\url{http://crocus.physics.mcmaster.ca/Cepheid/}} maintained by
Doug Welch. We thank Christiaan Sterken for communicating RV data by
\citet{2001JAD.....7....4C}. 

  This research is based on observations made with the Mercator Telescope,
  operated on the island of La Palma by the Flemish Community, at the Spanish Observatorio del Roque de
  los Muchachos of the Instituto de Astrof\'isica de Canarias.
  {\it HERMES} supported by the Fund for Scientific Research of
  Flanders (FWO), Belgium, the Research Council of K.U.~Leuven, Belgium, the Fonds National de
  la Recherche Scientifique (F.R.S.-FNRS), Belgium, the Royal Observatory of Belgium, the
  Observatoire de Gen\`eve, Switzerland, and the Th\"uringer Landessternwarte,
  Tautenburg, Germany. 
  
  This research has made use of NASA's ADS Bibliographic Services; the SIMBAD
  database and the VizieR catalogue access
  tool\footnote{\url{http://cdsweb.u-strasbg.fr/}} provided by CDS, Strasbourg;
  Astropy, a community-developed core Python package for Astronomy
  \citep{2013A+A...558A..33A}; the International Variable Star Index (VSX)
  database, operated at AAVSO, Cambridge, Massachusetts, USA.
   
RIA acknowledges funding from the Swiss National Science Foundation through an
Early Postdoc.Mobility fellowship. CM acknowledges support from the US National
Science Foundation through award NSF-AST-1313428.
PIP is a Postdoctoral Fellow of the The Research Foundation \hbox{--} Flanders
(FWO), Belgium. This research was supported by the Munich Institute for Astro-
and Particle Physics (MIAPP) of the DFG cluster of excellence "Origin and
Structure of the Universe".

\bibliographystyle{aasjournal} 
\bibliography{Bib_mine,Bib_Cepheids,Bib_modulation,Bib_interferometry,Bib_DistanceScale,Bib_Spectroscopy,ThesisBibTexRefs,RotatingCepBibTexRefs,Bib_StellarEvolution,Bib_Spectropolarimetry,Bib_general}

{\it Facilities:} \facility{Mercator1.2m}, \facility{Shane3m},
\facility{Euler1.2m}

\end{document}